\documentclass[prd,floatfix,nofootinbib,showpacs,showkeys]{revtex4}
\usepackage{exscale}                  
\usepackage[intlimits]{amsmath}       
\usepackage{amsfonts}
\usepackage{amssymb,amscd}
\usepackage[dvips]{epsfig}                   
\usepackage{array}
\usepackage{wrapfig}
\newcommand{\beqa}{\begin{eqnarray}}
\newcommand{\eeqa}{\end{eqnarray}}
\newcommand{\beq}{\begin{equation}}
\newcommand{\eeq}{\end{equation}}
\newcommand{\gS}[1]{#1\!\!\!\!\!\not~}	
\newcommand{\GS}[1]{#1\!\!\!\!\!\!\!\not~}	

\newcommand{\qslash}{\gS{q}}
\newcommand{\pslash}{p\!\cdot\!\gamma}
\newcommand{\Pslash}{\GS{P}}
\newcommand{\intk}{\int\!\!\frac{d^4k}{(2 \pi)^4}}

\newcommand{\intq}{\int\!\!\frac{d^4q}{(2 \pi)^4}}

\begin{document}

\large

\title{Hadronic unquenching effects in the quark propagator}
\author{Christian~S.~Fischer}
\author{Dominik~Nickel}
\author{Jochen~Wambach}
\affiliation{Institute for Nuclear Physics, Darmstadt University of
  Technology, Schlossgartenstra{\ss}e 9, 64289 Darmstadt, Germany}
\date{\today}

\begin{abstract}
We investigate hadronic unquenching effects in light quarks and mesons.
Within the non-perturbative continuum framework of Schwinger-Dyson and 
Bethe-Salpeter equations we quantify the strength of the back reaction 
of the pion onto the quark-gluon interaction. To this end we add a
Yang-Mills part of the interaction such that unquenched lattice results
for various current quark masses are reproduced. We find considerable 
effects in the quark mass function at low momenta as well as for the 
chiral condensate. The quark wave function is less affected. The 
Gell--Mann-Oakes-Renner relation is valid to good accuracy up to pion 
masses of 400-500 MeV. As a byproduct of our investigation we verify 
the Coleman theorem, that chiral symmetry cannot be broken spontaneously 
when QCD is reduced to 1+1 dimensions. 
\end{abstract}


\maketitle


\section{Introduction}\label{sec:intro}

Dynamical chiral symmetry breaking is one of the most important properties of
low-energy QCD. The breaking pattern has profound impact for phenomenological
quantities, as {\it e.g.} the appearance of the pseudoscalar Goldstone bosons 
in the chiral limit of QCD and the non-degeneracy of chiral partners.
Chiral perturbation theory~\cite{Weinberg:1978kz,Gasser:1983yg} describes these 
effects very efficiently  on the level of hadrons but has nothing to say about 
the underlying structure of the full theory. In this work we investigate 
dynamical chiral symmetry breaking on the level of (non-perturbative) correlation 
functions between quarks and gluons.

We are particularly interested in the interplay between the fundamental quark
and gluon degrees of freedom and the resulting bound states. In full QCD there
are hadronic contributions to the fully dressed quark-gluon interaction. These
effects are generated by the inclusion of dynamical sea quarks in the 
quark-gluon interaction and are therefore only present in unquenched QCD. In 
this work we concentrate on the back reaction of the pions on the quark
propagator and investigate the impact of the unquenched interaction on the
quark-pion system.

Pion effects on the quark propagation are important for several reasons. They 
account for (at least part of the) pion cloud effects in baryons and mesons. 
Thus they need to be incorporated in bound state calculations aiming at an 
adequate description of phenomenological properties of these
objects. Furthermore 
they allow for the possibility of hadronic intermediate states in bound state 
calculations and therefore generate the finite width of meson spectral
functions.
On a more fundamental level Gribov argued that pion effects in the quark 
propagator may be responsible for quark confinement~\cite{Gribov:1992tr}. 
In addition, the inclusion of these effects in a continuum framework
allows for  an extrapolation of unquenched lattice results for large
quark masses towards the physical up/down quark values. 
Finally, it will become apparent that Goldstone effects are essential
in smaller dimensions. In two dimensions they prohibit the dynamical
breaking of a continuous symmetry~\cite{Coleman:1973ci}.

We study the pion back reaction on the quarks in the Green's function approach 
to Landau gauge QCD using Schwinger-Dyson and Bethe-Salpeter equations
(SDE/BSE)~\cite{Maris:1997hd,Alkofer:2000wg,Maris:2003vk,Fischer:2006ub}. In the 
past  these effects have been investigated to some extent within the 
NJL-model~\cite{Dmitrasinovic:1995cb,Nikolov:1996jj,Oertel:2000jp}. Here we are
extending these model studies to full QCD. Unquenching effects in the quark-antiquark 
system have also been investigated by Watson and Cassing within the SDE/BSE approach 
in~\cite{Watson:2004jq}. There a coupled system of equations for the Bethe-Salpeter 
kernel and the connected quark four-point function has been solved in a certain 
approximation, which neglects the back reaction of the hadronic  resonances on the 
quark propagator. Our study here is
therefore in some sense complementary with their work. It is also complementary
to the investigations reported in~\cite{Fischer:2003rp,Fischer:2005en}, where 
unquenching effects in the gluon polarization have been considered.

In order to make our results as concise as possible we use a combination of 
two methods. We devise a truncation scheme for the combined Schwinger-Dyson 
equations for the gluon, ghost and quark propagator as well as the quark-gluon 
vertex, which is combined with the corresponding Bethe-Salpeter equation for 
light mesons. The details of the quark-gluon interaction are fixed such that 
the available unquenched lattice results for the quark propagator~\cite{Bowman:2005vx} 
are reproduced. This combination of methods 
allows us to study the relative impact of the pion contributions compared to 
the pure Yang-Mills part of the quark-gluon interaction.

The paper is organized as follows. In subsection \ref{sec:dse} we discuss the
hadronic contributions in the quark-gluon vertex and detail our approximation
scheme of this vertex and the quark Schwinger-Dyson equation (SDE). In
subsection  \ref{sec:kernel} we explain the extraction of pion
contributions in the  quark-antiquark scattering kernel of the
corresponding Bethe-Salpeter equation  (BSE). The resulting
interaction is supplemented by the familiar one-gluon 
interaction from the rainbow-ladder approximation. Both types of interaction 
together represent an approximation to the full unquenched quark-gluon interaction, 
which retains important properties of the mechanism of dynamical chiral symmetry 
breaking. A different perspective to this approximation in the framework of
2PI effective actions is discussed in subsection \ref{sec:Nc}. In subsection 
\ref{sec:fit} we explain our procedure to fit the parameters of the one-gluon 
exchange part of the interaction such that results from unquenched lattice 
simulations at large quark masses are recovered. Some technical details of our 
numerical method are presented in \ref{sec:numerics}. In subsection 
\ref{sec:quark-pion} we discuss our numerical results for the quark propagator 
and the pion at physical quark masses. We then investigate the chiral limit in 
subsection \ref{sec:chiral}. We show that the Gell--Mann-Oakes-Renner relation 
is satisfied in our approach and give a value for the chiral condensate. 
In subsection \ref{sec:analytical} a preliminary analysis for the analytical 
structure of the resulting quark-propagator is presented. Our results are 
summarized and discussed further in section \ref{sec:sum}. Throughout the paper 
we work in Euclidean space and in the isospin symmetric limit of equal up and 
down quark masses.

\section{The quark-gluon interaction}\label{sec:inter}

As emphasized above there is a distinct imprint of dynamical chiral symmetry
breaking on the low-energy meson spectrum: the lowest-lying pseudoscalar meson 
states are identified with the (pseudo-)Goldstone bosons of this mechanism. 
In the chiral limit, these states are exactly massless although they can also 
be described as bound states of constituent quarks with masses of the order
of 300-400 MeV. This dichotonuous nature of the pseudoscalar states has been 
discussed from first principles in~\cite{Maris:1997hd}. It is the axialvector 
Ward-Takahashi identity (axWTI), relating the quark self energy and the 
quark-antiquark scattering kernel, that enforces a binding energy of the 
pseudoscalar meson system which exactly cancels the quark and antiquark masses. 
It is therefore mandatory for every meaningful computational scheme of QCD to 
respect the axWTI. 

A widely used practical truncation of the SDE/BSE framework with this property 
is the rainbow-ladder approximation for the quark interaction. It has a history 
of remarkable successes (summarized {\it e.g.} in~\cite{Maris:2003vk}). However, 
there are also shortcomings that limit the credibility of such an approximation. 
Consequently, several efforts have been made to extend this scheme, see 
{\it e.g.}~\cite{Watson:2004kd,Bhagwat:2004kj,Bhagwat:2004hn,Fischer:2004ym,
Fischer:2005en,Maris:2005tt}. In the following we will extend the rainbow-ladder 
scheme by taking into account additional hadronic contributions to the quark-gluon 
interaction.

\subsection{SDEs for the quark-gluon vertex and the quark propagator}\label{sec:dse}

Let us first briefly sketch the so-called renormalization-group-improved
rainbow-ladder truncation. The starting point is the SDE of the quark propagator
\begin{eqnarray}
  S^{-1}(p) &=& Z_{2}S^{-1}_{0}(p)+\Sigma(p)\,, \label{DSE1}
\end{eqnarray}
where $S^{-1}_{0}(p) = i \pslash + m$ denotes the inverse bare quark-propagator, 
while $S^{-1}(p) = i \pslash A(p^2) + B(p^2)$ is the dressed propagator and $Z_{2}$ 
the renormalization factor of the quark field. The quark self energies $A(p^2)$ and 
$B(p^2)$ can be recombined into the quark mass $M(p^2) = B(p^2)/A(p^2)$ and the
 quark wave function $Z_f(p^2) = 1/A(p^2)$. The quark self energy is given by 
\begin{eqnarray}
  \Sigma(p) &=& g^{2}C_{F}Z_{1F}\intq\,
  \gamma_{\mu}S(q)\Gamma_{\nu}(q,k)
  D_{\mu\nu}(k)\,,
  \label{fullSigma}
\end{eqnarray}
with $k=p-q$, the Casimir $C_{F}=(N_{c}^{2}-1)/(2N_{c})$ and the renormalization 
factor $Z_{1F}$ of the quark gluon vertex. The self energy depends on the fully 
dressed quark-gluon vertex $\Gamma_{\nu}(q,k)$ and the gluon propagator 
\beq
D_{\mu\nu}(k) = \left(\delta_{\mu \nu} - \frac{k_\mu k_\nu}{k^2}\right) 
                                       \frac{Z(k^2)}{k^2}\,, \label{gluon}
\eeq
with the gluon dressing function $Z(k^2)$. Throughout this paper we will work in 
the Landau gauge, which guarantees the transversality of (\ref{gluon}). The 
rainbow-ladder approximation amounts to the replacement
\beq
\gamma_{\mu} Z(k^2) \Gamma_{\nu}(q,k) \rightarrow \gamma_{\mu} \Gamma(k^2) 
                                                  \gamma_{\nu}\,, \label{RL}
\eeq
where $\Gamma(k^2)$ can be viewed as a combination of the gluon dressing 
function and a purely $k^2$-dependent dressing of the $\gamma_\nu$-part of the
quark-gluon vertex. 

\begin{figure}[t]
\centerline{\epsfig{file=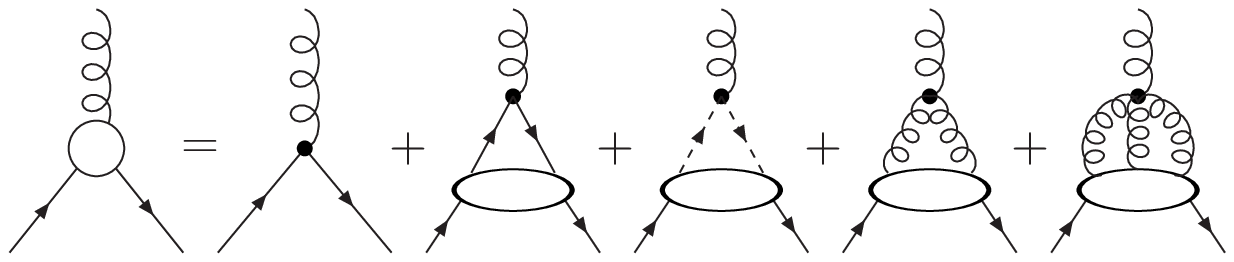,width=15cm}}
\vspace*{5mm}
\centerline{\epsfig{file=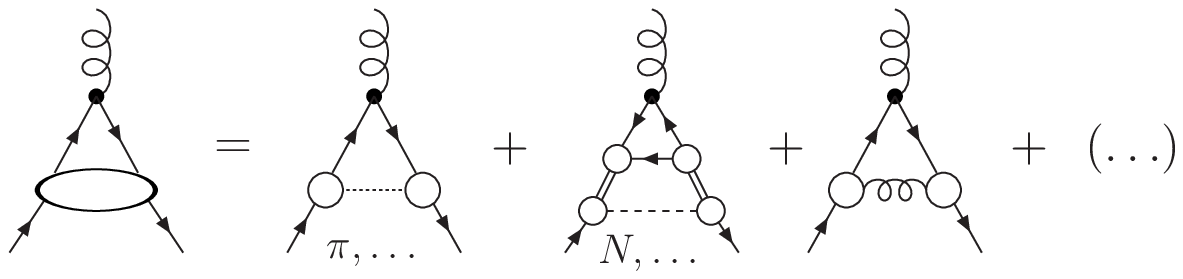,width=13cm}}
\caption{The full, untruncated Schwinger-Dyson equation for the quark-gluon
vertex \cite{Marciano:1977su} is shown diagrammatically in the first line.
The second line describes the first terms of an expansion in terms of hadronic 
and non-hadronic contributions to the quark-antiquark scattering kernel. In both 
equations, all internal propagators are fully dressed. Internal dashed lines with 
arrows correspond to ghost propagators, curly lines to gluons and full lines to 
quark propagators. In the second equation, the dotted line describes mesons, the 
dashed line baryons and the double lines correspond to diquarks.}
\label{fig:Vertexdse}
\end{figure}

Often the function $\Gamma(k^2)$ is referred to as an effective running 
coupling. Indeed it can be shown, that the ultraviolet behavior of 
$\Gamma(k^2)$ has to resemble that of the strong running coupling in order to 
reproduce the correct ultraviolet behavior of the quark propagator as known 
from resummed perturbation theory and the operator product 
expansion ~\cite{Politzer:1976tv}. The rainbow approximation then gives reliable 
results for the quark propagator in the ultraviolet. In the infrared, however, 
one needs to go beyond simple approximations like (\ref{RL}). The infrared shape 
of the quark propagator and also its analytical structure do depend on the details 
of the quark-gluon 
interaction~\cite{Fischer:2003rp,Alkofer:2003jj}. In particular, tensor
structures other than $\gamma_\nu$ seem to be important 
there~\cite{Alkofer:2006gz}. 

The Schwinger-Dyson equation for the quark-gluon vertex is given 
diagrammatically in Fig.~\ref{fig:Vertexdse}. In the first line we show the 
full, untruncated equation. For very small momenta, a selfconsistent solution 
to this equation has been given in ref.~\cite{Alkofer:2006gz}. Here we are 
primarily interested in the mid-momentum behavior of the vertex and in 
particular in hadronic contributions. To lowest order in a skeleton expansion 
such contributions can only occur in the diagram with the bare quark-gluon 
vertex at the external gluon line. In the second line of 
Fig.~\ref{fig:Vertexdse} we expand the quark-antiquark scattering kernel of this 
diagram in terms of resonance contributions to the kernel and one-particle 
irreducible Green's functions ('skeleton expansion'). Amongst other terms one 
obtains one-meson and one-gluon exchange between the quark and anti-quark lines.
Also diquark exchange contributions arise which are not explicitly shown in 
Fig.~\ref{fig:Vertexdse}. The first baryon exchange diagram shows up as a 
'two-loop' diagram, which also involves diquarks and the baryon-quark-diquark 
vertices. Note that double counting is trivially avoided in this combined 
expansion due to different quantum numbers in the exchange channel.

The computation of the hadronic diagrams in Fig.~\ref{fig:Vertexdse} is rather
involved. The meson-exchange diagram requires in addition the solution of a 
coupled system of the Schwinger-Dyson equation for the quark propagator and a 
corresponding Bethe-Salpeter equation (BSE) for the meson-quark vertex. Even 
more complex is the baryon-exchange diagram, which involves the computation of
the diquark-BSE and a Faddeev-type equation for the baryon bound state. In this
work we therefore wish to concentrate solely on contributions from the one-pion
exchange diagram in addition to the non-hadronic contributions. In fact this 
choice should be regarded as a first approximation to the fully unquenched system,
since all other hadronic diagrams are at least suppressed by factors of 
$\Lambda_{QCD}^2/m_H^2$ with $H \in \{K,\rho,N,...\}$. The same is true for diquark
exchange contributions.

On the quark level, the one-pion exchange diagram in Fig.~\ref{fig:Vertexdse}
involves a closed quark loop and therefore can only appear in unquenched QCD. 
Our investigation therefore complements previous studies of 
unquenching effects in the SDE framework, where quark-loops in the gluon 
polarization have been investigated~\cite{Fischer:2003rp,Fischer:2005en}. 

\begin{figure}[t]
\centerline{\epsfig{file=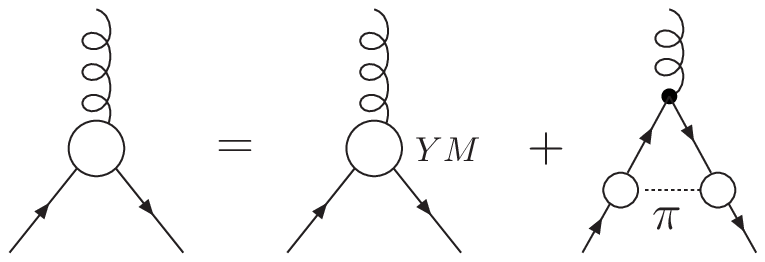,width=9cm}}
\caption{The approximated Schwinger-Dyson equation for the quark-gluon
vertex. All internal propagators are fully dressed. \label{fig:Vertexdse2}}
\vspace*{5mm}
\centerline{\epsfig{file=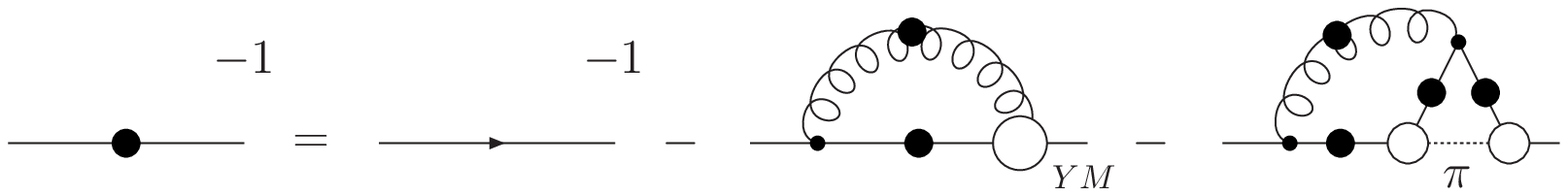,width=15cm}}
\caption{The Schwinger-Dyson equation for the quark propagator
with the quark-gluon vertex from Fig.~\ref{fig:Vertexdse2}.
\label{fig:quarkdse}}
\end{figure}

The Yang-Mills part of the vertex is conveniently approximated by 
\beq
\Gamma_\nu^{YM}(p_1,p_2,p_3) \;=\; \gamma_\nu\, \Gamma^{YM}(p_3^2) \,, \label{v1}
\eeq
where we denote the quark momenta by $p_1$ and $p_2$ and the gluon momentum by 
$p_3$. The explicit expression for $\Gamma^{YM}(p_3^2)$ is detailed in 
subsection \ref{sec:fit}. The ansatz (\ref{v1}) is an example for the
rainbow-ladder approximation, {\it cf.} Eq.~(\ref{RL}). A similar form has been
used in~\cite{Bhagwat:2003vw,Fischer:2005nf}, where quenched lattice results for the 
quark propagator have been reproduced. Note, however, that (\ref{v1}) 
involves only the $\gamma_\nu$-part of the tensor structure of the full vertex. 
 
In general one can decompose the quark-gluon vertex $\Gamma^\nu(p_1,p_2,p_3)$ 
into twelve different tensor structures, given by
\beq
\Gamma_\nu(p_1,p_2,p_3)=\sum_{i=1}^{12} \lambda_i(p_1,p_2,p_3) L^i_\nu (p_1,p_2,p_3)\,.
\eeq
The details of this basis are given in ref.~\cite{Skullerud:2002ge}. Here we
only note that $L^1_\nu = \gamma_{\nu}$; the explicit forms of the other 
structures are not needed. In principle, all twelve tensor structures can
be important in the intermediate momentum regime. This has been explored to some 
extent in lattice simulations \cite{Skullerud:2003qu}. We have verified by 
projection methods that the one-pion exchange diagram of
Fig.~\ref{fig:Vertexdse} contributes to all these structures.

These considerations lead to the following approximation scheme for the
quark-gluon vertex: we subsume the Yang-Mills part of the quark-gluon
interaction into a form as given in Eq.~(\ref{v1}) and add the one-pion 
exchange diagram of Fig.~\ref{fig:Vertexdse}. The resulting vertex is given 
in Fig.~\ref{fig:Vertexdse2}. It contains pion contributions in all twelve 
dressing functions $\lambda_i$ and contributions from the Yang-Mills sector in 
$\lambda_1$. A conceptually different justification of this approximation scheme 
in terms of a $1/N_c$-expansion is given in subsection \ref{sec:Nc}. Certainly, 
an explicit diagrammatic calculation of the Yang-Mills part of the vertex would 
be preferable to an ansatz of the form of 
Eq.~(\ref{v1}). Such a calculation is under way and will be detailed elsewhere.

Inserting the vertex of Fig.~\ref{fig:Vertexdse2} into the Schwinger-Dyson 
equation of the quark-propagator we arrive at the diagrammatic equation given 
in Fig.~\ref{fig:quarkdse}. In order to be able to construct a corresponding 
kernel for the Bethe-Salpeter equation of the pion we need to perform an 
additional approximation as detailed in the next subsection.

\subsection{The quark-antiquark scattering kernel}\label{sec:kernel}

We start with the general expression for the homogeneous Bethe-Salpeter equation 
(BSE) for quark-antiquark bound states, which can be written as
\beq
\Gamma_{tu}(p;P)=\int \frac{d^4k}{(2\pi)^4} K_{tu;rs}(p,k;P)
\left[S(k_+)\Gamma(k;P)S(k_-)\right]_{sr}\,.
\label{eq:bse}
\eeq
Here $\Gamma$ is the Bethe-Salpeter vertex function of a quark-antiquark
bound state and $K$ is the Bethe-Salpeter kernel. The momenta $k_+=k+\xi P$ and 
$k_-=k+(\xi-1)P$ are such that the total momentum is given by $P=k_+-k_-$.
The momentum partitioning parameter $\xi$ reflects the arbitrariness in
the relative momenta of the quark-antiquark pair and can be set to $\xi=1/2$ 
without loss of generality. Latin indices ($t,u,r,s$) refer to color, flavor 
and Dirac structure. The BSE is a parametric eigenvalue equation with discrete 
solutions $P^2=-M_n^2$ where $M_n$ is the mass of the resonance. The lowest 
mass solution corresponds to the physical ground state. Since $P^2$ is 
negative, the momenta $k_{\pm}$ are necessarily complex in Euclidean space 
and thus the quark propagator functions are evaluated with complex argument. We 
will come back to this issue below.

\begin{figure}[t]
\centerline{\epsfig{file=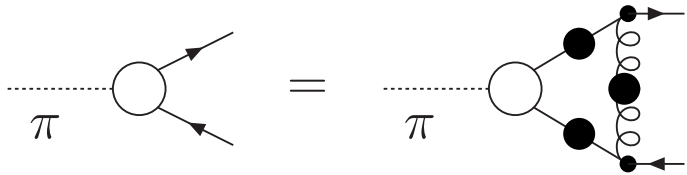,width=10cm}}
\caption{The Bethe-Salpeter equation in rainbow-ladder approximation, 
{\it {\it i.e.}} effective one-gluon exchange. 
\label{fig:kernelYM}}
\vspace*{5mm}
\end{figure}

Chiral symmetry constrains the Bethe-Salpeter kernel $K_{tu;rs}$ via the 
axialvector Ward-Takahashi identity (axWTI),
\beq
\left[\Sigma(p_+)\gamma_5 + \gamma_5\Sigma(p_-)\right]_{tu}
=\int \frac{d^4k}{(2\pi)^4} K_{tu;sr}(p,k;P) \left[\gamma_5 S(k_-)+S(k_+)
                                                      \gamma_5\right]_{rs},
\label{eq:axwti}
\eeq
which relates the kernel to the quark self energy 
$\Sigma(p) = S^{-1}(p)-Z_{2}S^{-1}_{0}(p)$, {\it cf.} Eq.~(\ref{DSE1}). In the 
rainbow-ladder approximation this relation can be satisfied easily. For our 
Yang-Mills part of the interaction, Eq.~(\ref{v1}), one then obtains the kernel
\beq
  K_{tu;rs}^{YM}(q,p;P)
  =
  \frac{g^2 \, Z(k^2)\,  \Gamma^{YM}(k^2) \, Z_{1F}}{k^{2}}
  \left(\delta_{\mu\nu}-\frac{k_{\mu}k_{\nu}}{k^{2}}\right)
  \left[\frac{\lambda^{a}}{2}\gamma_{\mu}\right]_{ts}
  \left[\frac{\lambda^{a}}{2}\gamma_{\nu}\right]_{ru}\,. \label{YMkernel}
\eeq
The resulting BSE includes an effective one-gluon exchange between the 
quark-antiquark pair as shown in Fig.~\ref{fig:kernelYM}. 

The construction of a Bethe-Salpeter kernel corresponding to the pion-exchange
part of our interaction is more complicated. We were not able to find an exact 
solution of the axWTI for the interaction given in Fig.~\ref{fig:Vertexdse2}. 
In principle one then has two choices: either one works with an approximate 
solution of the axWTI or one approximates the interaction further such that an 
exact solution of the axWTI to the approximated interaction is possible. Since 
the first option destroys the Goldstone character of the pion, only the second
choice is viable and shall be followed here.

\begin{figure}[t]
\centerline{\epsfig{file=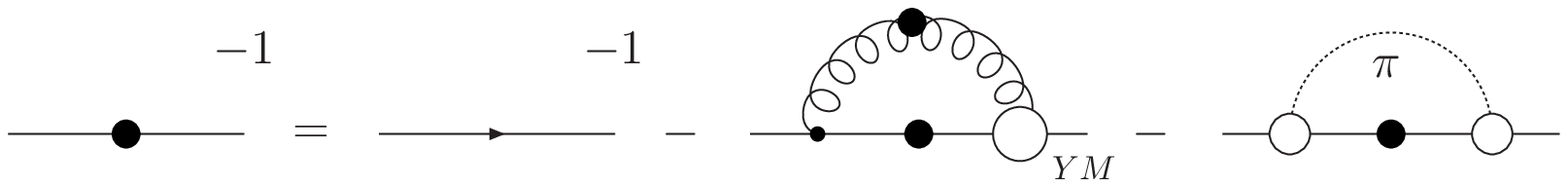,width=16cm}}
\caption{The approximated Schwinger-Dyson equation for the quark propagator
with effective one-gluon exchange and one-pion exchange. 
\label{fig:quarkdse2}}
\vspace*{10mm}
\centerline{\epsfig{file=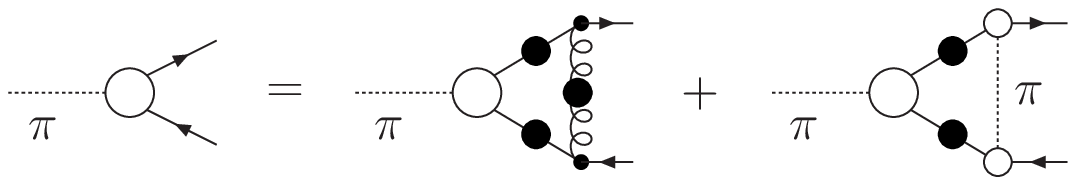,width=15cm}}
\caption{The Bethe-Salpeter equation corresponding to the quark self energy
of Fig.~\ref{fig:quarkdse2}.
\label{fig:BSE}}
\vspace*{5mm}
\end{figure}

To this end we reconsider the quark-SDE given in Fig.~\ref{fig:quarkdse}. For 
diagrammatical reasons the corresponding Bethe-Salpeter kernel should
look like 
a superposition of the effective one-gluon exchange and an interaction
involving pions. Let us assume for the moment that only the Yang-Mills
part of the interaction would be present in the BSE. Then the
quark-SDE could be rewritten by inserting the BSE into the second
diagram on the right hand side. The result is given diagrammatically
in Fig.~\ref{fig:quarkdse2} and represents an interaction given by an
effective one-gluon and a one-pion exchange diagram in the SDE.

The only remaining question concerns the description of the one-pion
exchange, which will be needed in the $t$-channel (see
Fig.~\ref{fig:quarkdse} and Fig.~\ref{fig:quarkdse2}).
In the $s$-channel the pion contribution to the quark-antiquark
scattering kernel is given by~\cite{Maris:1997hd}
\begin{eqnarray}
  \label{eq:schannel}
  M_{tu;rs}^{pion}(q,p;P)
  &=&
  [\bar{\Gamma}^i_{\pi}]_{rs}\left(q;-P\right)
  \frac{1}{P^{2}+M_{\pi}^{2}}
  [\Gamma^j_{\pi}]_{tu}\left(p;P\right)
  +
  O\left((P^{2}+M_{\pi}^{2})^{0}\right)
  \,,
\end{eqnarray}
where $\Gamma^i_{\pi}$ is the Bethe-Salpeter vertex function,
$\bar{\Gamma}^i_{\pi}\left(q;-P\right)=
C^{-1}\Gamma^{i}_{\pi}\left(-q;-P\right)^{T}C$ and
$C=\gamma_{2}\gamma_{4}$.
As already stated we neglect other contributions in the quark-antiquark
scattering kernel as we expect the Goldstone modes to be dominant.
The corresponding expression for the $t$-channel should be related by
crossing symmetry (if we take incoming and outgoing particles
on-shell). As the latter will however be broken when only considering the
leading term in Eq.~(\ref{eq:schannel}), a transformation of the
leading term in Eq.~(\ref{eq:schannel}) to the $t$-channel is not
unique and depends on the choice for incoming and outgoing momentum.
In order to find a suited kernel for the BSE which fulfills the axWTI
in Eq.~(\ref{eq:axwti}) later on, we take the arithmetic mean of the two
possible contributions and obtain for the self-energy
\begin{eqnarray}
  S^{-1}(p) &=& Z_2 \,S^{-1}_0(p) + g^2 \, C_F \, Z_{1F} 
  \int \frac{d^4q}{(2\pi)^4} \, \gamma_\mu \, S(q) \, \gamma_\nu \, 
  \left(\delta_{\mu\nu}-\frac{k_\mu k_\nu}{k^2}\right) \frac{Z(k^2) 
    \Gamma_{YM}(k^2)}{k^2} \nonumber\\
  &&-3 \int \frac{d^4q}{(2\pi)^4} \left[      
    \Gamma_{\pi}\left(\frac{p+q}{2};p-q\right)\, S(q)\,
    \Gamma_{\pi}\left(\frac{p+q}{2};q-p\right) \right.\nonumber\\
    &&\hspace*{21mm}+\left.
    \Gamma_{\pi}\left(\frac{p+q}{2};q-p\right)\, S(q)\,
    \Gamma_{\pi}\left(\frac{p+q}{2};p-q\right)
    \right] \frac{D_{\pi}(k^2)}{2}
\end{eqnarray}
with $k=p-q$ and $D_{\pi}(k)=(k^{2}+M_{\pi}^{2})^{-1}$ being the pion
propagator.
The factor $3$ in front of the pion contribution stems from the 
flavor trace and represents contributions from $\pi_+, \pi_-$ and $\pi_0$. 
These are treated on equal footing in the isospin-symmetric limit
adopted here.
The general structure of the 
Bethe-Salpeter vertex of the pion can be represented by
\beq
\Gamma^i_{\pi}(p,P) = \tau^i \gamma_5 \left\{ E_\pi(p,P) - i \Pslash\, F_\pi(p,P) 
                    - i \pslash \,p \cdot P\, G_\pi(p,P) 
                    - [\Pslash,\pslash] \,H_\pi(p,P) \right\} \label{pion}
\eeq
with four independent dressing functions $E_\pi,F_\pi,G_\pi$ and $H_\pi$. 
Those obey $f(k;P)=f(-k;-P)=f(-k;P)$ for
$f\in\{E_{\pi},F_{\pi},G_{\pi},H_{\pi}\}$.
For the case $H_{\pi}=0$ it is then possible to respect the axWTI in
Eq.~(\ref{eq:axwti}) for $P^{2}\rightarrow 0$ in the chiral limit by
choosing the kernel
\begin{eqnarray}
  \label{Kapprox1}
  K_{tu;rs}^{pion}(q,p;P)
  &=&
  -\frac{3}{2}
      [\Gamma^j_{\pi}]_{ts}\left( \frac{p+q+P}{2};p-q\right)
      [\Gamma^j_{\pi}]_{ru}\left(-\frac{p+q+P}{2};p-q\right)
       D_{\pi}(p-q)
  \nonumber\\
  &&-\frac{3}{2}
      [\Gamma^j_{\pi}]_{ts}\left( \frac{p+q+P}{2};q-p\right)
      [\Gamma^j_{\pi}]_{ru}\left(-\frac{p+q+P}{2};q-p\right)
      D_{\pi}(p-q)    \,.\phantom{ccc}
\end{eqnarray}
Neglecting $H_{\pi}$ would therefore be a suited approximation,
especially since it is known to have very small impact on hadronic observables
(although the vertex function is then, strictly speaking, no longer
Poincare invariant).
We will however follow another strategy here by symmetrizing our
kernel in $(q,p,P)\leftrightarrow (-q,-p,P)$ for which we obtain
\begin{eqnarray}
  \label{Kapprox}
  K_{tu;rs}^{pion}(q,p;P)
  &=&
  -\frac{3}{4}
      [\Gamma^j_{\pi}]_{ts}\left( \frac{p+q+P}{2};p-q\right)
      [\Gamma^j_{\pi}]_{ru}\left(-\frac{p+q+P}{2};p-q\right)
       D_{\pi}(p-q)
  \nonumber\\
  &&-\frac{3}{4}
      [\Gamma^j_{\pi}]_{ts}\left( \frac{p+q+P}{2};q-p\right)
      [\Gamma^j_{\pi}]_{ru}\left(-\frac{p+q+P}{2};q-p\right)
      D_{\pi}(p-q)    \nonumber\\
  &&-\frac{3}{4}
      [\Gamma^j_{\pi}]_{ts}\left(-\frac{p+q-P}{2};p-q\right)
      [\Gamma^j_{\pi}]_{ru}\left( \frac{p+q-P}{2};p-q\right)
       D_{\pi}(p-q)
  \nonumber\\
  &&-\frac{3}{4}
      [\Gamma^j_{\pi}]_{ts}\left(-\frac{p+q-P}{2};q-p\right)
      [\Gamma^j_{\pi}]_{ru}\left( \frac{p+q-P}{2};q-p\right)
      D_{\pi}(p-q)\,.    \nonumber\\
\end{eqnarray}
This is the only kernel we have found to analytically respect the
axWTI for $P^{2}\rightarrow 0$ in the chiral limit for the general
structure of the Bethe-Salpeter vertex given in Eq.~(\ref{pion}).
From a theoretical viewpoint it is less well grounded than the
choice Eq.~(\ref{Kapprox1}): whereas (\ref{Kapprox1}) can be represented
by proper Feynman diagrams the symmetrized version Eq.~(\ref{Kapprox})
does not share this property. It is however motivated from a pragmatic 
point of view since it allows one to check the influence of 
$H_{\pi}$ explicitly. We will therefore prefer the version Eq.~(\ref{Kapprox1})
in this work. As a result we will see that neglecting contributions 
from $H_{\pi}$ introduces errors on the level of a view percent into
observables such as the pion decay constant. For many applications it 
may therefore be possible to use the simpler kernel (\ref{Kapprox1}).
Both kernels have the correct charge conjugation properties and respect
multiplicative renormalizability of the BSE by construction.

The resulting diagrammatic expression of the Bethe-Salpeter equation
is shown in  Fig.~\ref{fig:BSE}. The sum over all contribution in
Eq.~(\ref{Kapprox}) is presented by the one-pion exchange diagram.
Our approximation of the quark-DSE in
Fig.~(\ref{fig:quarkdse}) by the one given in
Fig.~(\ref{fig:quarkdse2}) is justified provided the Bethe-Salpeter
vertex functions of the pion with or without the pion interaction term
do not differ strongly. We will show in section \ref{sec:quark-pion}
that  this is indeed the case.

With the kernel (\ref{Kapprox}) our final expression for the quark-SDE is then 
given by
\beqa
S^{-1}(p) &=& Z_2 \,S^{-1}_0(p) + g^2 \, C_F \, Z_{1F} 
\int \frac{d^4q}{(2\pi)^4} \, \gamma_\mu \, S(q) \, \gamma_\nu \, 
\left(\delta_{\mu\nu}-\frac{k_\mu k_\nu}{k^2}\right) \frac{Z(k^2) 
\Gamma_{YM}(k^2)}{k^2} \nonumber\\
&&- 3 \int \frac{d^4q}{(2\pi)^4} \left[      
      \Gamma_{\pi}\left(\phantom{-}\frac{p+q}{2};p-q\right)\, S(q)\,
      \Gamma_{\pi}\left(\phantom{-} \frac{p+q}{2};q-p\right) \right.\nonumber\\
      &&\hspace*{21mm}+
      \Gamma_{\pi}\left(\phantom{-} \frac{p+q}{2};q-p\right)\, S(q)\,
      \Gamma_{\pi}\left(\phantom{-} \frac{p+q}{2};p-q\right) \nonumber\\ 
      &&\hspace*{21mm}+
      \Gamma_{\pi}\left(-\frac{p+q}{2};p-q\right)\, S(q)\,
      \Gamma_{\pi}\left(-\frac{p+q}{2};q-p\right)  \nonumber\\
      &&\hspace*{21mm}\left.+
      \Gamma_{\pi}\left(-\frac{p+q}{2};q-p\right)\, S(q)\,
      \Gamma_{\pi}\left(-\frac{p+q}{2};p-q\right) \right] \frac{D_{\pi}(k^2)}{4} 
      \label{quarkdse2}
\eeqa
with $k=p-q$.
The corresponding expression for the BSE reads
\beq
\Gamma_{tu}(p;P)=\int \frac{d^4k}{(2\pi)^4} \left\{K_{tu;rs}^{YM}(p,k;P) 
                     + K_{tu;rs}^{pion}(p,k;P)\right\}
                       \left[S(k_+)\Gamma(k;P)S(k_-)\right]_{sr}
\label{eq:bse2}
\eeq
with the kernels $K_{tu;rs}^{YM}$ and $K_{tu;rs}^{pion}$ given in 
Eqs.~(\ref{YMkernel}) and (\ref{Kapprox}). The 
Bethe-Salpeter vertex of the pion, i.e. the dressing functions
$E_\pi,F_\pi,G_\pi$ and $H_\pi$ are normalized according to the condition 
\beqa
2\delta^{ij}P_{\mu} &=& 3 \, \, Tr_d\int \frac{d^4k}{(2\pi)^4} \left\{
\overline{\Gamma}_\pi^i(k,-P) \frac{\partial S(k+P/2)}{\partial P_{\mu}}
          \Gamma_\pi^j(k,P)S(k-P/2) \right.\nonumber\\
&& \hspace*{2cm}\left.
+\overline{\Gamma}_\pi^i(k,-P)S(k+P/2)\Gamma_\pi^j(k,P)\frac{\partial S(k-P/2)}
{\partial P_{\mu}}\right\} \nonumber\\
&& + \int \frac{d^4k}{(2\pi)^4}\int \frac{d^4q}{(2\pi)^4} 
[\overline{\chi}_\pi^i]_{sr}(q,-P)
\frac{\partial K_{tu;rs}^{pion}(q,k;P)}{\partial P_\mu}
[{\chi}_\pi^j]_{ut}(k,P)
\label{norm} 
\eeqa
with ${\chi}_\pi^j(k,P) = S(q+P/2) \Gamma_\pi^j(k,P) S(q-P/2)$. The condition
is written for the momentum partitioning $\xi=1/2$ without loss of generality. 
The trace is over Dirac matrices and the conjugate vertex function 
$\overline{\Gamma}$ is defined as
\beq
\overline{\Gamma}(p,-P)=C\Gamma^T(-p,-P)C^{-1}
\eeq
with the charge conjugation matrix $C=-\gamma_2\gamma_4$. The normalization 
condition (\ref{norm}) guarantees that the corresponding residue of the pion 
pole in the four-point quark-antiquark Green's function is 
unity~\cite{Nakanishi:1969ph}. The leptonic decay constant characterizing 
the pion coupling to the point axial field is then given by~\cite{Tandy:1997qf}
\beq
f_{\pi}=\left.\frac{Z_2 N_c}{M_\pi^2}\, \, Tr_d\int \frac{d^4k}{(2\pi)^4}
\Gamma_{\pi}(k,-P)S(k+P/2)\gamma_5 \, \Pslash \, S(k-P/2)\right|_{P^{2}=-M_{\pi}^{2}}\,,
\label{eq:fpi}
\eeq
where again the trace is over Dirac matrices. The numerical details needed for 
solving Eqs.~(\ref{quarkdse2}) and (\ref{eq:bse2}) simultaneously are
discussed in subsection \ref{sec:numerics}. We wish to emphasize that the 
approximations made in the course of this section respect multiplicative 
renormalizability. This is shown explicitly in appendix \ref{MR}. Since 
also the axWTI is preserved we may hope that our approximation scheme 
captures essential properties of full QCD.

In this context there are some interesting points to note:
\begin{itemize}
\item A crudely simplified version of our interaction was considered 
by Gribov in his investigations of quark confinement due to 
supercritical charges \cite{Gribov:1992tr}. He conjectured that the 
inclusion of the pion back reaction to the quark induces important 
changes in the analytic structure of the quark propagator. We will 
come back to this point in subsection \ref{sec:analytical}.
\item We observe the quark self energy in
Eq.~(\ref{quarkdse2}) to be infrared divergent in two dimensions for
any non-vanishing Bethe-Salpeter amplitude in the chiral limit.
More precisely the axWTI requires $\Gamma^{i}_{\pi}(p,0)=\tau^{i}\gamma_5
B(p)/f_{\pi}$~\cite{Maris:1997hd} and we have in good approximation
\beqa
\Sigma^{pion}(p)
&\simeq&
-3\int\!\!\frac{d^dq}{(2 \pi)^d}
\left(\frac{B(\frac{p+q}{2})}{f_{\pi}}\right)^{2}
\frac{-i\qslash A(q^{2})+B(q^{2})}{q^{2}A^{2}(q^{2})+B(q^{2})}
\frac{1}{k^{2}}
\eeqa
for $d$ dimensions.
For any finite value of the dressing function $B(p^{2})$ at any
momentum $p$, the integral of the pion contribution is infrared
singular for $d=2$ and would
shift the value $B(p^{2})\rightarrow-\infty$,
{\it i.e.} would give an infinite repulsive back reaction.
Thus only the trivial solution $B(p^2)=0$ is left in two dimensions.
This supports Coleman's theorem~\cite{Coleman:1973ci},
which states that continuous symmetries in two dimensions cannot be
broken dynamically.
\item
Using the approximation $\Gamma^{i}_{\pi}(p,P)\approx\tau^{i}\gamma_5
B(p)/f_{\pi}$~\cite{Maris:1997hd} in the $s$-channel contribution of
the pion to
the quark-antiquark scattering kernel (see Eq.~(\ref{eq:schannel})) a
simpler choice for $K_{tu;rs}^{pion}(q,p;P)$ could be
\begin{eqnarray}
  \label{eq:Kpionsimple}
  K_{tu;rs}^{pion}(q,p;P)
  &=&
  -3[\tau^j \gamma_5\frac{B(\frac{p+q}{2})}{f_{\pi}}]_{ts}
  \,D_{\pi}(p-q)\,
  [\tau^j \gamma_5\frac{B(\frac{p+q}{2})}{f_{\pi}}]_{ru}
  \,.
\end{eqnarray}
We consider this as an approximation to our construction for the kernel.
\item  
Finally we note that the inclusion of a back reaction by Goldstone
bosons is possible for any case of dynamical symmetry breaking on an
equal footing, {\it
  e.g.} it can similarly be applied to color-superconducting phases
elaborated in~\cite{Nickel:2006vf} for our approach.
\end{itemize}
In the following subsection we will reconsider our approximation
scheme from a different perspective.

\subsection{The $1/N_c$-expansion}\label{sec:Nc}

\begin{figure}
  \begin{center}
    \includegraphics[width=4.cm]{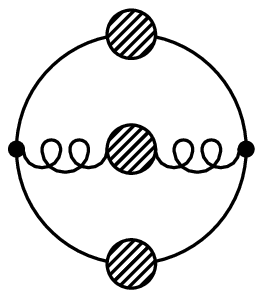}
    \hspace{2cm}
    \includegraphics[width=4.cm]{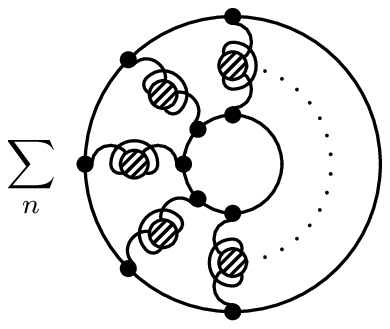}
  \end{center}
  \caption{
    Leading (sunset diagram on the left) and next-to-leading (right, the sum
    is over the number of gluon lines) order contribution to $\Gamma_{2}[S]$
    in a model with effective one-gluon exchange. All signs and prefactors
    have been absorbed in the diagrams.
  }
  \label{fig:Nc}
\end{figure}

The approximation specified in the last subsection may also be viewed in the 
light of the $1/N_c$-expansion. The pion exchange contributions are then
the (approximate) next-to-leading order correction to the leading one-gluon
exchange of the pure Yang-Mills interaction. Such a view has been advocated
already in model frameworks such as the 
NJL-model~\cite{Dmitrasinovic:1995cb,Nikolov:1996jj,Oertel:2000jp}.
In the following we discuss this aspect also for QCD in the framework
of two-particle irreducible (2PI)
effective actions, {\it i.e.} the CJT-formalism, which also allows for a 
derivation of the SDEs for propagators.
The CJT action~\cite{Cornwall:1974vz} for truncations with effective
one-gluon exchange is a 
functional of the quark propagator and given by
\begin{eqnarray}
  \label{eq:CJT}
  \Gamma[S] &=& 
  -\mathrm{Tr}\mathrm{Ln}\left(Z_{2}^{-1}S_{0}S^{-1}\right)
  + \mathrm{Tr}
  \left(1-Z_{2}S_{0}^{-1}S\right)+
  \Gamma_{2}[S]
  \,.
\end{eqnarray}
Here we already neglected the dependence on the expectation value of quark
fields $\langle\psi\rangle$ and $\langle\bar{\psi}\rangle$ as they are
vanishing for the ground state.

For the diagrammatic expansion of the 2PI functional
$\Gamma_{2}[S]$ we can then use a $1/N_{c}$-ordering.
The $1/N_{c}$-expansion of vacuum-vacuum contributions needed for
$\Gamma_{2}[S]$ can nicely be arranged by the topology of the contributing
diagrams~\cite{'tHooft:1973jz}.
Since we do not consider pure gluonic contributions at order $O(N_{c}^{2})$,
the leading contribution is given by planar gluonic diagrams with a quark line
as a boundary, being $O(N_{c})$.
The corresponding contribution to $\Gamma_{2}[S]$ is the left diagram
of Fig.~\ref{fig:Nc}.
To next-to-leading order, {\it i.e.} $O(1)$ , we have the topology of a cylinder
with two quark lines as boundaries.
Those contributions to $\Gamma_{2}[S]$ are subsumed in the right diagram
of Fig.~\ref{fig:Nc}.

Therefore the truncated quark-SDE to next-to-leading order in a
$1/N_{c}$-expansion, given by $\frac{\delta\Gamma[S]}{\delta S}=0$, turns out
to be
\begin{eqnarray}
  S^{-1}(p) &=& Z_{2}S^{-1}_{0}(p)+\Sigma^{(0)}(p)+\Sigma^{(1)}(p)
  \,,
\end{eqnarray}
where $\Sigma^{0}(p)=\Sigma^{(rainbow)}(p)$ is the rainbow self energy
from Eqs.~(\ref{fullSigma}-\ref{RL}) and
\begin{eqnarray}
  \label{eq:sig2}
  \Sigma^{(1)}_{ts}(p) &=&
  -\intq\,
  S_{ur}(q)\,M_{tu;rs}\left(\frac{p+q}{2},\frac{p+q}{2};p-q\right)
\end{eqnarray}
with
\begin{eqnarray}
  \label{eq:Mladder2}
  M_{tu;rs}(q,p;P) &=&
  K^{YM}_{tu;rs}(q,p;P)
  \nonumber\\&&+
  \intk\,
  K^{YM}_{tu;vw}(q,k;P)S_{wa}(k_{+})
  M_{ab;rs}(k,p;P)S_{bv}(k_{-})
\end{eqnarray}
is the next-to-leading order contribution.

Since we do not want to perform the ladder resummation
Eq.~(\ref{eq:Mladder2}), we furthermore want to approximate
$M_{tu;rs}(q,p;P)$ in order to reduce the numerical complexity.
First we note that Eq.~(\ref{eq:sig2}) is still correct up to
next-to-leading order in $1/N_{c}$ when using the propagators from the
rainbow-ladder approximation.
The matrix $M_{tu;rs}(q,p;P)$ then contains Goldstone bosons in the
$s$-channel. As we again presume the Goldstone bosons to be dominant, we will
focus on the leading term in Eq.~(\ref{eq:schannel}).
Following our discussion how to obtain the pion contribution to the
quark self-energy given in Eq.~(\ref{quarkdse2}) from the $s$-channel
contribution in Eq.~(\ref{eq:schannel}), we can similarly motivate the
same result for the self-energy as given in Eq.~(\ref{eq:sig2}):
When using Eq.(\ref{eq:schannel}) in Eq.~(\ref{eq:sig2}), we are
unable to find a suited kernel for the Bethe-Salpeter equation which
respects chiral symmetry.
The axWTI then suggests to use
\begin{eqnarray}
  \label{eq:MNcApprox}
  M_{tu;rs}(q,p;P)
  &=&
  \frac{1}{4}
       [\Gamma^{a}_{\pi}]_{tu}(\phantom{-}q;\phantom{-}P)
  \frac{1}{P^{2}+M_{\pi}^{2}}
       [\Gamma^{a}_{\pi}]_{rs}(\phantom{-}p;-P)
  +
  \nonumber\\&&
  \frac{1}{4}
       [\Gamma^{a}_{\pi}]_{tu}(\phantom{-}q;-P)
  \frac{1}{P^{2}+M_{\pi}^{2}}
       [\Gamma^{a}_{\pi}]_{rs}(\phantom{-}p;\phantom{-}P)
  +
  \nonumber\\&&
  \frac{1}{4}
       [\Gamma^{a}_{\pi}]_{tu}(-q;\phantom{-}P)
  \frac{1}{P^{2}+M_{\pi}^{2}}
       [\Gamma^{a}_{\pi}]_{rs}(-p;-P)
  +
  \nonumber\\&&
  \frac{1}{4}
       [\Gamma^{a}_{\pi}]_{tu}(-q;-P)
  \frac{1}{P^{2}+M_{\pi}^{2}}
       [\Gamma^{a}_{\pi}]_{rs}(-p;\phantom{-}P)
\end{eqnarray}
for the ladder resummed fully-amputated quark-antiquark scattering
amplitude.
This is again a pragmatic approach as we need to symmetrize the
arguments in $M_{tu;rs}(q,p;P)$ in order to find a manageable
Bethe-Salpeter kernel.
We then recover the quark-SDE in Eq.~(\ref{quarkdse2}) as the approximate
next-to-leading order expression within the CJT-formalism.

Strictly speaking we would then need to use the rainbow-ladder
Bethe-Salpeter vertex functions in the expression for the self-energy,
however to be consistent up to next-to-leading order in $1/N_{c}$ we
can also insert next-to leading order expressions.
A simpler truncation would be to use the approximation given in
Eq.~(\ref{eq:Kpionsimple}).

The Bethe-Salpeter kernel is then obtained by an additional functional
derivative~\cite{Munczek:1994zz}
\begin{eqnarray}
  K_{tu;rs}
  &=&
  -\frac{\delta^{2}\Sigma_{tu}}{\delta S_{rs}}\,.
  \label{eq:Sigmavaried}
\end{eqnarray}
Using our approximated next-to-leading order contribution of the self energy
and neglecting the dependence of the Bethe-Salpeter amplitudes on the propagators 
then allows to justify the kernel given in Eq.~(\ref{Kapprox}).
Essentially the kernel guarantees the pion to be massless in the chiral limit.

\begin{figure}[t]
\centerline{\epsfig{file=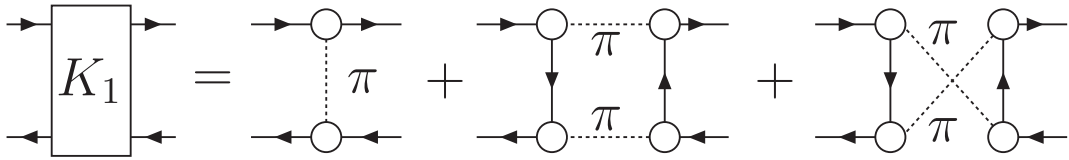,width=15cm}}
\caption{The Bethe-Salpeter kernel $K_1$ including next to leading order 
contributions in the pion interaction, {\it cf.}
Eq.~(\ref{eq:KNc}). The dotted line in this context constitutes the
resummed ladder being introduced in the $1/N_{c}$-expansion. In
particular it includes the pseudoscalar channel.
\label{fig:extkernel}}
\end{figure}

Taking the $1/N_{c}$-expansion more serious by performing the ladder summation
results in a more complicated expression for the Bethe-Salpeter kernel. We
need to evaluate the derivative in Eq.~(\ref{eq:Sigmavaried}). 
Also including the momentum in the multi-indices we abbreviate
Eq.~(\ref{eq:sig2}) by
\begin{eqnarray}
  \Sigma^{(1)}_{ts}
  &=&
  -S_{ur}M_{tu;rs}
  \,.
\end{eqnarray}
We then get
\begin{eqnarray}
  K^{(1)}_{tu;rs}
  &=&
  M_{ts;ru}+
  S_{xy}
  \frac{
    \delta
  }{
    \delta S_{rs}
  }
  M_{tx;yu}
\end{eqnarray}
for the next-to-leading order contribution to the
Bethe-Salpeter kernel.
The leading order contribution is simply given by the effective
one-gluon exchange.
Evaluating the derivative by using Eq.~(\ref{eq:Mladder2}) yields
\begin{eqnarray}
  \frac{
    \delta
  }{
    \delta S_{rs}
  }
  M_{tx;yu}
  &=&
  K_{tx;ms}M_{rb;yu}S_{bm}+
  K_{tx;rn}M_{as;yu}S_{na}+
  \nonumber\\&&
  K_{tx;ma}
  S_{ma}
  S_{bm}
  \frac{
    \delta
  }{
    \delta S_{rs}
  }
  M_{ab;yu}
  \,.
\end{eqnarray}
Now we can use Eq.~(\ref{eq:Mladder2}) to resum the right hand side of this equation.
We then end up with
\begin{eqnarray}
  \frac{
    \delta
  }{
    \delta S_{rs}
  }
  M_{tx;yu}
  &=&
  M_{tx;ms}M_{rb;yu}S_{bm}+
  M_{tx;rn}M_{as;yu}S_{na}
  \,.
\end{eqnarray}
Summarizing this analysis we obtain the next-to-leading order 
contribution to the Bethe-Salpeter kernel to be of the form
\begin{eqnarray}
  K^{(1)}_{tu;rs}
  &=&
  M_{ts;ur}
  +
  M_{tx;ms}M_{rb;yu}S_{bm}S_{xy}
  +
  M_{tx;rn}M_{as;yu}S_{na}S_{xy}
  \,,
  \label{eq:KNc}
\end{eqnarray}
which corresponds to a resummed ladder in the $t$-channel and two
contributions with two resummed ladders in the $s$-channel.
The contributing diagrams are actually similar to those 
introduced in~\cite{Watson:2004jq}. 
Having discussed these terms we can think of Eq.~(\ref{eq:MNcApprox}) as an
approximation of the first term of Eq.~(\ref{eq:KNc}), which is
identical to the one introduced in subsection \ref{sec:kernel}.
A diagrammatic presentation of Eq.~(\ref{eq:KNc}) is shown in Fig.~\ref{fig:extkernel}.

\subsection{Gluonic interaction}\label{sec:fit}

As discussed in previous subsections we need to specify two different 
components of the Yang-Mills part of the interaction: the gluon dressing 
function $Z(p^2)$ and the vertex dressing $\Gamma_{YM}$. The gluon 
dressing function has been calculated numerically from a truncated 
version of the Schwinger-Dyson equations for the ghost and gluon 
propagator~\cite{Fischer:2002hn,Fischer:2003rp}. The resulting dressing 
functions for the ghost and gluon propagator have been discussed and 
compared with corresponding lattice results in previous works, see
{\it {\it e.g.}}~\cite{Fischer:2007pf} and references therein. In order 
to make this paper self-contained we use functional forms for the ghost 
and gluon dressing functions which represent the numerical solutions to 
sufficient accuracy~\cite{Alkofer:2003jj}. These forms are given by
\beqa
  Z(p^2) &=&   \left(\frac{p^2}{\Lambda^2_{YM} + p^2}\right)^{2 \kappa} 
               \left(\frac{\alpha_{\rm fit}(p^2)}
	       {\alpha_{\rm fit}(\mu^2)}\right)^{-\gamma}\;, \nonumber\\
  G(p^2) &=&   \left(\frac{p^2}{\Lambda^2_{YM} + p^2}\right)^{-\kappa} 
               \left(\frac{\alpha_{\rm fit}(p^2)}
	       {\alpha_{\rm fit}(\mu^2)}\right)^{-\delta}\;,
  \label{fullfit}  
\eeqa 
with the running coupling
\begin{equation}
\alpha_{\rm fit}(p^2) =  
\frac{\alpha(0)}{1+ p^2/\Lambda^2_{YM}}
 + \frac{4 \pi}{\beta_0} \; \frac{p^2}{p^2 + \Lambda^2_{YM}} \;
      \left(\frac{1}{\ln(p^2/\Lambda^2_{YM})}- 
            \frac{\Lambda_{YM}^2}{p^2-\Lambda_{YM}^2}\right) \; ,
\label{fitB}
\end{equation}
and the renormalization condition $\alpha(\mu^2) = 0.968$. The scale 
$\Lambda_{YM}=0.510 \,\mbox{GeV}$ is scheme dependent, here it corresponds 
to the momentum subtraction (MOM) scheme used 
in~\cite{Fischer:2002hn,Fischer:2003rp}. Via the analytic structure
(\ref{fullfit}) of the gluon dressing function it can be related to
the 
physical energy scale where positivity violations in the gluon propagator 
occur~\cite{Alkofer:2003jj}. 
The infrared exponent $\kappa$ has been determined in an analytical infrared 
analysis~\cite{Zwanziger:2001kw,Lerche:2002ep}: 
$\kappa = (93-\sqrt{1201})/98 \approx 0.595$. The ultraviolet anomalous 
dimension of the ghost is given by $\delta=-9N_c/(44N_c-8N_f)$ and related via 
$\gamma=-1-2\delta$ to the one of the gluon.
Furthermore we have $\beta_{0}=4/\gamma_{m}=(11N_{c}-2N_{f})/3$.
Here we use $N_c=3$ and $N_f=2$.
Unquenching effects due to quark-loops in the gluon polarization are included 
in the expressions (\ref{fullfit}). 

\begin{figure}[t]
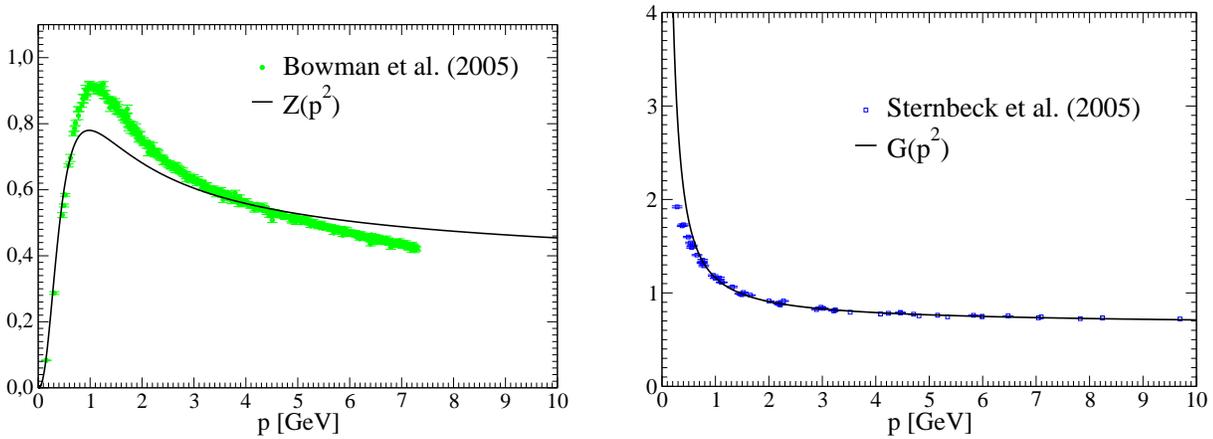

\centerline{\epsfig{file=lattglue.eps,width=7.5cm}\hfill
            \epsfig{file=lattghost.eps,width=7.5cm}}
\caption{SDE solutions for the gluon dressing function $Z(p^2)$ and the ghost 
dressing function $G(p^2)$ adapted from ref.~\cite{Fischer:2002hn} and compared 
to the lattice results of~\cite{Bowman:2004jm,Sternbeck:2005tk}.
\label{fig:ghostglue}}
\end{figure}

We show (\ref{fullfit}) together with lattice results for the unquenched 
($N_f=2+1$) gluon dressing function~\cite{Bowman:2004jm} and the 
quenched\footnote{Lattice calculations of the unquenched ghost dressing 
function are not yet available.} ghost dressing 
function~\cite{Sternbeck:2005tk} 
in Fig.~\ref{fig:ghostglue}. For the gluon both results are in good agreement 
in the infrared momentum region (deviations are discussed 
in~\cite{Fischer:2007pf}). In the intermediate momentum region truncation artifacts 
in the gluon-SDE lead to an underestimation of the hump in the dressing function. 
In the ultraviolet the SDE-solutions show the correct logarithmic scaling as 
expected from resummed perturbation theory, whereas the lattice data suffer 
from artifacts due to the finite lattice spacing. The ghost dressing function 
show slight deviations in the infrared, which are discussed in detail 
in~\cite{Fischer:2007pf}. In general we believe that both solutions are in 
sufficient agreement such that the expressions (\ref{fullfit}) serve as a 
reliable input for the quark SDE.

For the second part of the Yang-Mills interaction, the vertex dressing 
$\Gamma_{YM}$, we use a procedure first suggested in~\cite{Bhagwat:2003vw} and 
further explored in~\cite{Fischer:2005nf}: we devise a functional form for the
vertex dressing with parameters fixed by the requirement that lattice data for
the quark propagator are reproduced by solutions of the quark-SDE using the 
complete interaction. The functional form we employ for this procedure
is given by
\begin{equation}
\Gamma_\nu(k,\mu^2) \;=\;\gamma_\nu \Gamma_{YM}(k,\mu^2) \;=\; 
\gamma_\nu\, \Gamma_{1}(k^2)\, \Gamma_{2}(k^2,\mu^2)\, \Gamma_{3}(k^2,\mu^2) 
\label{v2}
\end{equation}
with the components
\begin{eqnarray}
\Gamma_{1}(k^2) &=&  \frac{\pi \gamma_m}{\ln(k^2/\Lambda_{QCD}^2 +\tau)}\,, 
\label{v3}\\[3mm]
\Gamma_{2}(k^2,\mu^2) &=& \sqrt{\frac{k^2}{k^2 + \Lambda_{YM}^2}} G(k^2,\mu^2)
\ G(\zeta^2,\mu^2)\ \widetilde{Z}_3(\mu^2) 
 \ h \ [\ln(k^2/\Lambda_{YM}^2 +\tau)]^{1+\delta} \label{v4} \\[3mm]
\Gamma_{3}(k^2,\mu^2) &=& Z_2(\mu^2)\; \frac{a(M)+k^2/\Lambda_{QCD}^2}
{1+k^2/\Lambda_{QCD}^2}\,, \label{v5}
\end{eqnarray}
where $\tau = e-1$ acts as a convenient infrared cutoff for the
logarithms.
As mentioned in section \ref{sec:dse} other tensor structures in the
Yang-Mills part of the quark-gluon vertex might be relevant, an 
investigation of those is however beyond the scope of this work. 
We will come back to this point at the end of section \ref{sec:fit}.
In the following we discuss the choice of our ansatz (\ref{v2}) step by
step and compare it to the one used in the quenched calculation of
ref.~\cite{Fischer:2005nf}.

It is well known that the effective interaction in the quark-SDE has
to approach the running coupling in the ultraviolet momentum 
regime~\cite{Miransky:1985ib}, {\it i.e.} 
\beq
\frac{g^2}{4\pi} \frac{1}{Z_2 \widetilde{Z}_3}Z(k^2) \Gamma_{1}(k^2) 
    \Gamma_{2}(k^2) \Gamma_{3}(k^2)
 \rightarrow 
  \frac{\pi \gamma_m}{\ln(k^2/\Lambda_{QCD}^2)}. \label{UV} 
  \eeq
Up to constants this UV-behavior is represented by $\Gamma_1$ with the 
scheme-dependent scale $\Lambda_{QCD}$. The anomalous dimension $\gamma_m$ 
of the quark propagator is given by $\gamma_m = 12/(33 - 2 N_f)$, where we 
use $N_f=2$ in our calculations. 
Furthermore note that $Z_{2}=\widetilde{Z}_{3}Z_{1F}$ in Landau gauge, where
$\widetilde{Z}_{3}$ is the renormalization factor of the ghost fields.

The product $Z(k^2) \Gamma_{2}(k^2) \Gamma_{3}(k^2)$ goes to a constant for 
large momenta; the scale $\Lambda_{YM}$ is determined from the SDE-solutions 
for the ghost and gluon propagator (\ref{fullfit}). The coefficient $h$ is 
fixed such that the limit (\ref{UV}) is satisfied. The renormalization group 
invariant $G(\zeta^2,\mu^2)\widetilde{Z}_3(\mu^2)$ with the arbitrary but 
fixed scale $\zeta$ is introduced to impose the correct cutoff- and 
renormalization point dependencies of the effective interaction in the quark-SDE. 
In our calculations we use $\zeta=2.9$ GeV, other choices do not affect the result.

The infrared behavior of our interaction is dominated by $\Gamma_2$, which is 
diverging for $p^2 \rightarrow 0$ whereas $\Gamma_1$ and $\Gamma_3$ go to 
constants. The degree of divergence is given by 
$G(k^2) \sqrt{\frac{k^2}{k^2 + \Lambda_{YM}^2}} \sim (k^2)^{-\kappa-1/2}$. This 
behavior has been derived in ref.~\cite{Alkofer:2006gz} from an analytic, 
selfconsistent analysis of the full SDE for the quark-gluon vertex given in 
Fig.~\ref{fig:Vertexdse}. We wish to emphasize, however, that the precise 
infrared behavior of $\Gamma_{YM}$ is not important for the quark-SDE: this 
equation is dominated by intermediate loop momenta. One could equally well work 
with lower degrees of divergence~\cite{Fischer:2005nf} or even with an infrared 
finite vertex~\cite{Bhagwat:2003vw}. 

Finally, from the Slavnov-Taylor identity of the quark-gluon vertex one can see 
that the vertex is also proportional to the quark wave function $Z_f$. This 
dependence is taken care of by $\Gamma_3$, which mimics the momentum dependence 
of $Z_f$. The extra factor $Z_2$ is vital for ensuring multiplicative 
renormalizability of the quark-SDE. The dependence of this part of the vertex 
on the quark mass is expressed in terms of the function
\beq
a(M)\; =\; \frac{a_1}
   {1 + a_2 M(\zeta^2)/\Lambda_{QCD} + a_3 M^2(\zeta^2)/\Lambda_{QCD}^2}.
\label{v6}
\eeq
In order to preserve multiplicative renormalizability of the quark-SDE, it is 
important that the scale $\zeta$ is not correlated with the renormalization 
point. Instead it should be a fixed scale. In our calculations we use the same 
scale $\zeta=2.9$ GeV as also present in $\Gamma_2$.

\begin{figure}[t]
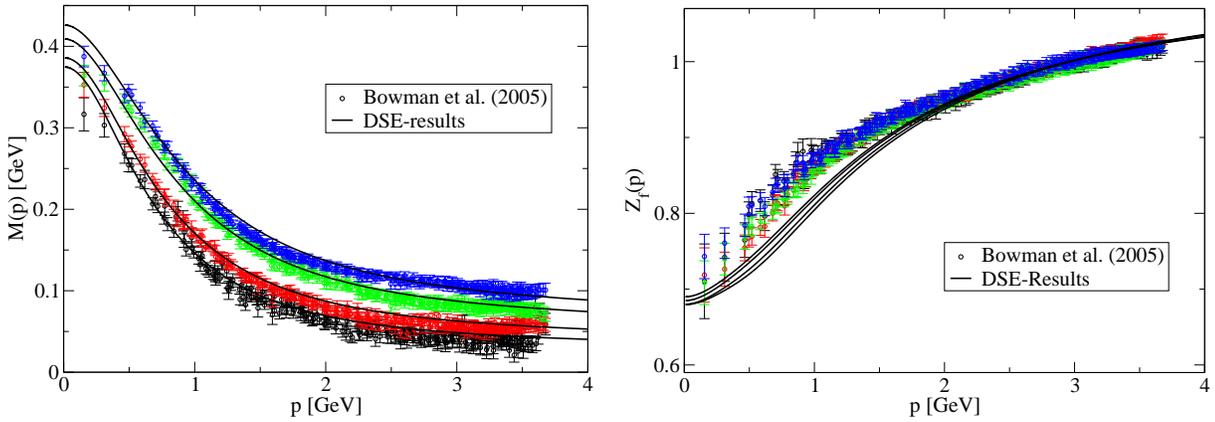

\centerline{\epsfig{file=Lattice_M.eps,width=7.8cm}
      \hfill\epsfig{file=Lattice_Z.eps,width=7.8cm}}
\caption{SDE results compared to lattice data of the unquenched quark
mass function $M(p)$ and the wave function $Z_f(p)$~\cite{Bowman:2005vx}.}
\label{fig:latt}
\vspace{0mm}
\end{figure}
\begin{table}
\begin{tabular}{|c|c|c|c|c|c|}
\hline
     \hspace*{4mm} h                    \hspace*{4mm} 
  & \hspace*{1mm} $\Lambda_{YM}$(GeV)  \hspace*{1mm} 
  & \hspace*{1mm} $\Lambda_{QCD}$(GeV) \hspace*{1mm} 
  & \hspace*{4mm} $a_1$                \hspace*{4mm}
  & \hspace*{4mm} $a_2$                \hspace*{4mm} 
  & \hspace*{4mm} $a_3$                \hspace*{4mm}\rule[-3mm]{0mm}{7mm}\\
  \hline\hline
             0.99(1) & 0.51(5)  & 0.65(5) & 5.22 (1) & 5.06 (1) & -9.06 (1) 
	     \rule[-3mm]{0mm}{7mm} \\ \hline
\end{tabular}
\vspace{3mm}
\caption{Parameters used in the vertex model, Eqs.~(\ref{v2}-\ref{v6}).
          \label{tab}}
\vspace{5mm}
\end{table}

To fit the parameters of the Yang-Mills part of our interaction we use the 
following procedure: We first determine the values of $h, \Lambda_{QCD}$ and 
$a(M)$ at a fixed current quark mass such that (i) the correct ultraviolet 
behavior (\ref{UV}) of the running coupling is reproduced and (ii) the 
unquenched lattice quark propagators from ref.~\cite{Bowman:2005vx} are 
reproduced. The Yang-Mills scale $\Lambda_{YM}$ is taken from the SDE-results 
parameterized in Eqs.~(\ref{fullfit}). This procedure is repeated for the four 
different current quark masses of ref.~\cite{Bowman:2005vx}, which correspond
to $M(2.9 \,\mbox{GeV})=44,65,85,106 \,\mbox{MeV}$.
We then fit Eq.~(\ref{v6}) to the results for $a(M)$ and determine $a_1, a_2$ 
and $a_3$. The results for the parameters are given in Tab.~\ref{tab}.  

The numerical results of our fitting procedure are shown in
Fig.~\ref{fig:latt}. For the quark mass function $M(p^2)$ we find
excellent agreement with the lattice data in both, the infrared and
ultraviolet momentum regions. For the wave function $Z_f(p^2)$ the
agreement is only slightly less convincing. Particularly interesting
is the small spread of the wave functions for different quark masses,
which is reproduced by the SDE results. This is in contrast to earlier
investigations of the quenched quark propagator, where the spread in
the SDE results has been too large~\cite{Bhagwat:2003vw,Fischer:2005nf}. We 
consider the improved result here as an indication that our quark-gluon 
interaction is more realistic than in previously used models. 
In particular we attribute this to the fact that the hadronic contributions
in the quark-gluon interaction generate all twelve possible Dirac structures
of this interaction (only eight of these are independent in Landau gauge),
whereas the quenched investigations \cite{Bhagwat:2003vw,Fischer:2005nf} 
mainly worked with the $\gamma_\mu$-structure only. This is a clear hint
that one should include further structures in the modeling of the Yang-Mills 
part $\Gamma_{YM}$ of the interaction and is left for future work.
Our solutions for the quark mass function $M(p^2)$ and the wave function 
$Z_f(p^2)$ for physical up/down quarks are discussed in section \ref{sec:quark-pion}.

\section{Numerical Procedure and Results for quark and pion}\label{sec:results}

\subsection{Numerical Procedure}\label{sec:numerics}

\begin{figure}[t]
\centerline{\epsfig{file=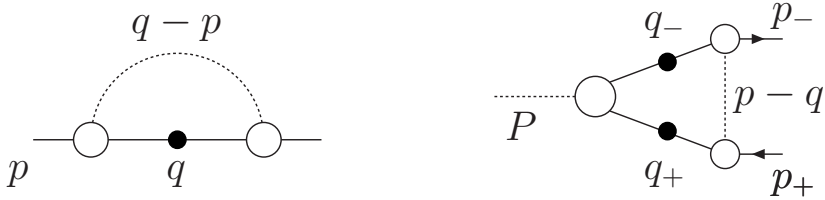,width=11cm}}
\caption{The momentum routing in the SDE and BSE.}
\label{fig:momentum}
\vspace{0mm}
\end{figure}

Before presenting results we need to detail our numerical procedures. Our 
choice of the momentum routing in the SDE and BSE is specified in
Fig.~\ref{fig:momentum}. In the right diagram $P$ denotes the total momentum 
of the meson and $q_+=q + P/2$, $q_-=q -P/2$ and $p_+=p + P/2$, $p_-=p -P/2$  
are the momenta of the internal and external quarks. Since in the rest frame 
of the meson we  have $P_\mu = (0,0,0,i M)$, where $M$ is the mass of the 
bound state, we need to know the quark propagators in the BSE for complex 
momenta. In addition, for the pion exchange diagram we need to know the pion 
Bethe-Salpeter amplitude for complex relative momenta. Whereas solving the 
quark-SDE in the complex plane is feasible and a standard procedure by now (see 
{\it e.g.}~\cite{Maris:1997tm,Alkofer:2002bp,Fischer:2005en}), the solution of
 the BSE for
complex momenta has not yet been performed to our knowledge. In view of the
considerable numerical complexity of the coupled system of SDE and BSE we will
not attempt such a solution within the context of this work. Instead we resort 
to the well explored 'absolute value approximation', which amounts to replacing 
all momentum arguments in internal quark propagators in the BSE by their 
absolute value. In addition, we replace the arguments of the pion amplitudes 
in the meson exchange diagrams by their real parts. This approximation has the 
merit of making the kernel independent of $P_\mu$, which in turn greatly 
simplifies the Bethe-Salpeter norm (\ref{norm}). We will attempt a complete 
solution of the problem in the complex plane in future work. 

To estimate the quality of this approximation we employed the (quenched) model 
of~\cite{Alkofer:2002bp} and solved it once for complex momenta and once in the 
absolute value approximation for a physical pion. Using similar current quark 
masses the resulting difference in the pion mass is less than 1 MeV. The error 
in $f_\pi$ however is larger: we obtain $f_\pi=94$ MeV with complex momenta 
but only $f_\pi = 82$ MeV in the absolute value approximation. Roughly $10$ MeV 
of this difference is due to the absolute value approximation in the norm 
integral (\ref{norm}); the remaining $2$ MeV are accounted for by shape 
distortions in the $F$, $G$ and $H$ amplitudes of the pion Bethe-Salpeter 
amplitude (\ref{pion}). We expect similar errors when using the quenched 
version of our interaction, {\it i.e.} only taking into account the gluon 
exchange part of the interaction given by Eq.~(\ref{v2}). In the unquenched 
calculations the error in the norm integral is backfeeding into the coupled 
system by the overall normalization of the pion contribution of the interaction. 
We therefore expect somewhat larger errors in both, the pion mass and the 
resulting pion decay constant. Nevertheless we consider our results to be 
accurate and meaningful in their qualitative features.

Further details concerning the numerical techniques needed to solve the 
quark-SDE are given in \cite{Fischer:2003rp,Fischer:2005en}; corresponding 
techniques for the BSE are described in~\cite{Maris:1997hd,Fischer:2005en}.

\subsection{Quark propagator and pion properties}\label{sec:quark-pion}

\begin{figure}[t]
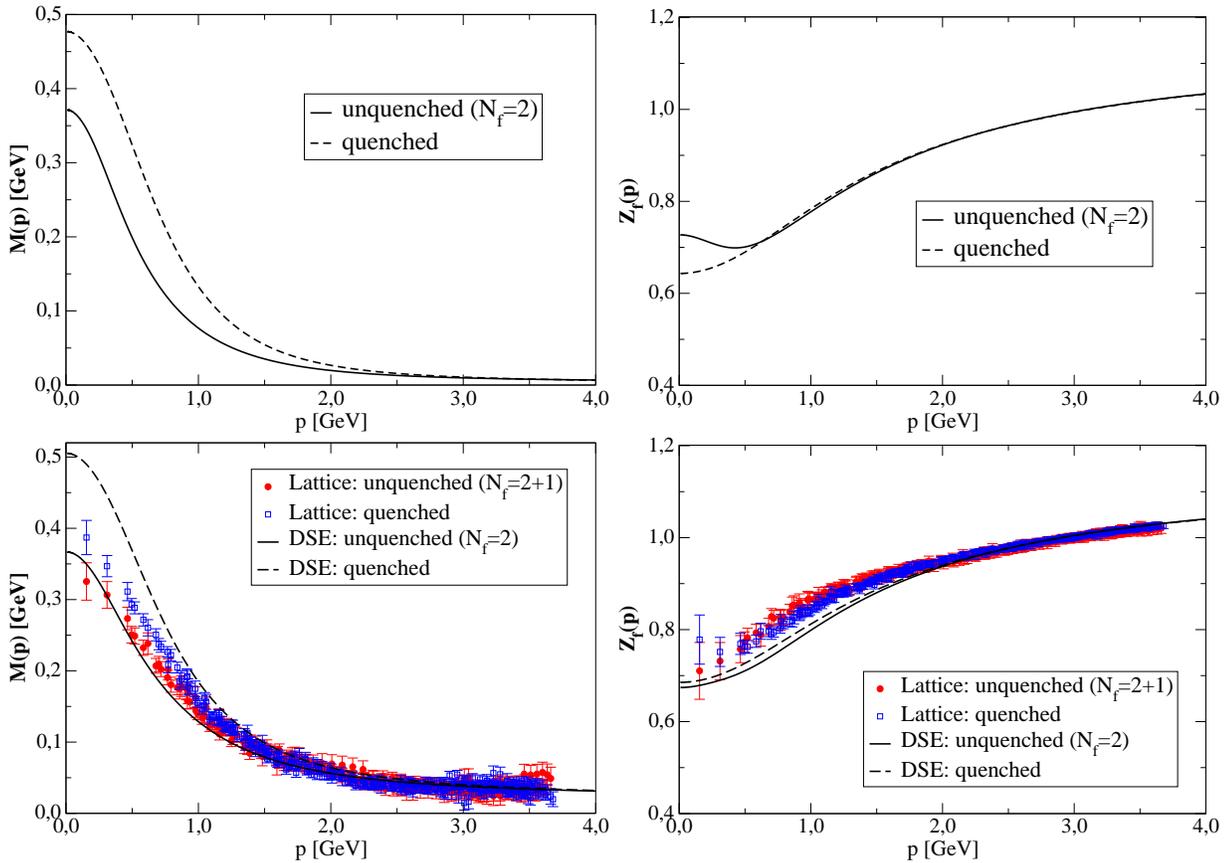

\centerline{\epsfig{file=PI_M.eps,width=8cm}
      \hfill\epsfig{file=PI_Z.eps,width=8cm}}
\centerline{\epsfig{file=Mvslatt.eps,width=8cm}
      \hfill\epsfig{file=Zvslatt.eps,width=8cm}}
\caption{In the upper panel we show the quenched and unquenched
  ($N_f=2$) quark mass function (left diagrams) and wavefunction (right diagrams) 
  for physical up/down quarks with $M(2.9 \,\mbox{GeV})=10 \,\mbox{MeV}$. In the 
  lower panel we compare the
  SDE/BSE-solutions for a heavier quark of $M(2.9 \,\mbox{GeV})=44 \,\mbox{MeV}$
  to the quenched and unquenched lattice results of Bowman et al. \cite{Bowman:2005vx}.}
\label{fig:quark}
\vspace{0mm}
\end{figure}

Our results for the physical up/down quark propagator are shown in
Fig.~\ref{fig:quark}. We compare the unquenched ($N_f=2$) solutions from the 
SDE/BSE system with partially quenched solutions. Here, 'partial quenching' 
means that we still include the quark loop effects in the gluon propagator, 
implicitly present in Eqs.~(\ref{fullfit}), but switch off the hadronic 
unquenching effects from the pion part of the interaction. In both calculations 
the renormalized current quark mass is chosen such that the pion mass is 
$M_{\pi} = 136$ MeV. In an $\overline{MS}$-scheme with $\Lambda_{\overline{MS}} = 225$ 
MeV this corresponds to renormalized up/down current quark masses of
\beq
m_{u/d}^{N_f=2}(2 \,\mbox{GeV}) = 3.5 \,\mbox{MeV} \hspace*{1cm} 
m_{u/d}^{N_f=0}(2 \,\mbox{GeV}) = 2.8 \,\mbox{MeV}.
\eeq
Note that the unquenched current quark mass is larger than the quenched one. In 
\cite{Fischer:2005en} a small effect in the opposite direction has been found 
from quark loops in the gluon polarization alone. The stronger increase found 
here reverses this tendency and indicates that unquenching effects in the 
quark-gluon vertex are stronger than those in the gluon polarization.

\begin{figure}[t]
\centerline{\epsfig{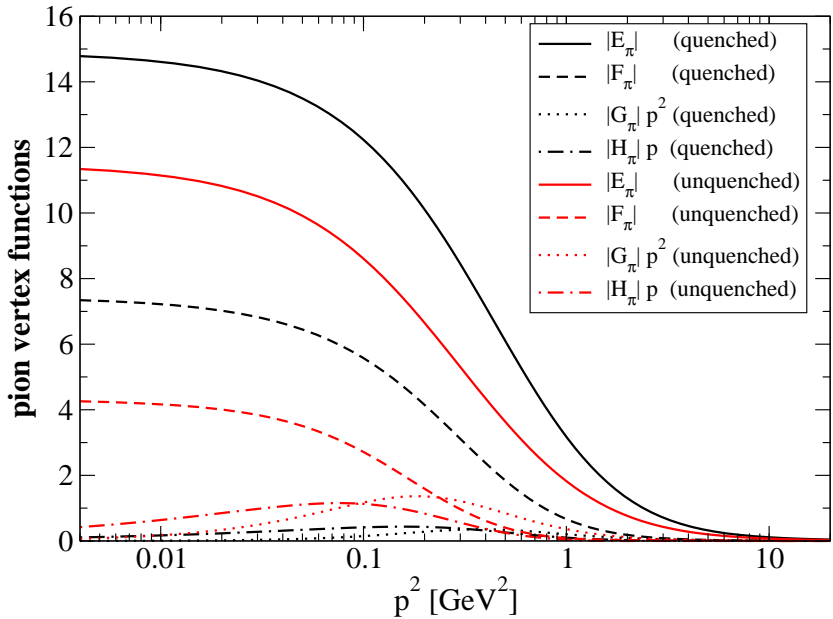}}
\caption{The quenched and unquenched ($N_f=2$) pion Bethe-Salpeter amplitudes.}
\label{fig:pi}
\vspace{0mm}
\end{figure}

The quark mass function in the upper left diagram of Fig.~\ref{fig:quark} also 
shows a sizable reduction of the dynamically generated mass $M(p^2)$ in the 
infrared when the theory is unquenched. These effect is greatly diminished at 
$p = 2$ GeV and vanishes for momenta larger than $p = 3$ GeV. The corresponding 
effect in the quark wave function $Z_f(p^2)$ is much smaller and only visible 
for momenta lower than $p= 0.5 $ GeV. This finding agrees qualitatively with 
results from quenched and unquenched ($N_f=2+1$) lattice calculations shown 
in the lower panel of Fig.~\ref{fig:quark}. Clearly, both approaches underpin 
the physical intuition: in general, quark loop effects are screening and 
compensate part of the antiscreening gluonic interaction. This leads to a 
reduction of the interaction strength in the gluon 
propagator~\cite{Fischer:2003rp,Bowman:2004jm}. Here it also reduces the 
strength of the quark-gluon vertex and consequently entails (further) reduced 
quark masses. 
The effects for the quark mass function seen in the SDE/BSE approach are  
larger than in the lattice calculations. This may or may not be an artifact of
the absolute value approximation used in the numerics. In any case, it seems 
save to conclude that unquenching effects from the pion back reaction ({\it i.e.} 
in the quark-gluon vertex) are considerably stronger than the quark-loop effects 
in the gluon polarization investigated in refs.~\cite{Fischer:2003rp,Fischer:2005en}.

The corresponding Bethe-Salpeter amplitudes of the pion are shown in
Fig.~\ref{fig:pi}. Similar to the previous findings in the quenched 
approximation~\cite{Maris:1997tm} we find a hierarchy in magnitude of these 
amplitudes: clearly $E_\pi$ is most important in size followed by $F_\pi$, 
$G_\pi$ and $H_\pi$. This hierarchy is not changed by unquenching, although 
we note a considerable increase in size of $G_\pi$ and $H_\pi$ for $N_f=2$ 
compared to the quenched case. Despite these differences both sets of amplitudes 
are similar enough to justify the approximation made in subsection \ref{sec:kernel}.
In the infrared, $E_\pi$, $F_\pi$, $G_\pi$ and $H_\pi$ go to 
constants. In the ultraviolet $E_\pi$ and $F_\pi$ fall like $1/p^2$ times a 
logarithm, whereas $G_\pi$ and $H_\pi$ are proportional to $1/p^4$ times a logarithm. 
This behavior is not changed when the system is unquenched. 
\begin{table}
\begin{center}
\begin{tabular}{|c||c|c|c|c|c|c|c|}
\hline
          & \hspace*{2mm} $M(0)$           \hspace*{2mm} 
          & \hspace*{2mm} $M_\pi$          \hspace*{2mm} 
	  & \hspace*{1mm} $f_\pi$ (full)   \hspace*{1mm} 
	  & \hspace*{1mm} $f_\pi$ from E   \hspace*{1mm} 
	  & \hspace*{2mm} $f_\pi$ from F   \hspace*{2mm}
	  & \hspace*{2mm} $f_\pi$ from G   \hspace*{2mm} 
	  & \hspace*{2mm} $f_\pi$ from H   \hspace*{2mm} 
	                                \rule[-3mm]{0mm}{7mm}\\ \hline\hline
E       & 329  & 131  & 40 & 40  & 0  & 0  & 0 \rule[-3mm]{0mm}{7mm} \\\hline
E,F     & 353  & 140  & 53 & 45  & 8  & 0  & 0 \rule[-3mm]{0mm}{7mm} \\\hline
E,F,G   & 369  & 145  & 43 & 45  & 6  & -8 & 0 \rule[-3mm]{0mm}{7mm} \\\hline
E,F,G,H & 371  & 136  & 49 & 46  & 7  & -9 & 6 \rule[-3mm]{0mm}{7mm} \\\hline
\end{tabular}
\vspace{3mm}
\caption{The influence of the four components of the Bethe-Salpeter amplitude
of the pion on masses and decay constants. All results are for the unquenched 
interaction; the units are in MeV.\label{tab:fpi}}
\end{center}
\vspace{5mm}
\end{table}
The relative importance of the Bethe-Salpeter amplitudes can be inferred from 
table \ref{tab:fpi}. We show results for the quark mass function at zero momentum, 
the pion mass and the contributions to the pion decay constant when part of the 
amplitudes are omitted. On a quantitative level it is certainly important to include 
all four amplitudes in the calculation. If, however, one is satisfied with qualitative 
results, even a calculation involving only the $E_\pi$-amplitude would give satisfactory 
results.

Using all four Bethe-Salpeter amplitudes, the resulting values for the pion decay 
constant calculated from (\ref{eq:fpi}) in the absolute value approximation and for 
a physical value of the current quark masses are
\beq
f_{\pi}^{N_f=2} = 49 \,\mbox{MeV} \hspace*{1cm} f_{\pi}^{N_f=0} = 82 \,\mbox{MeV}. 
     \label{fpi}
\eeq
Clearly, unquenching reduces the value of $f_\pi$. However, the magnitude of 
this reduction is too large. Also both values are too low. One would expect the
unquenched value close to the physical point $f_{\pi}^{exp} = 93$ MeV and a 
somewhat larger number in the quenched case. A reduction of the unquenched pion 
decay constant was also found in recent lattice calculations with 
quenched~\cite{Jansen:2005kk} and unquenched~\cite{Boucaud:2007uk} twisted mass 
fermions. They obtained
\beq
f_{\pi,lattice}^{N_f=2} = 86(1) \,\mbox{MeV} \hspace*{1cm} 
f_{\pi,lattice}^{N_f=0} = 102(3) \,\mbox{MeV}. \label{fpilattice}
\eeq
for our definition of the pion decay constant and in the chiral limit. In the 
unquenched simulation the pion decay constant at the physical point was used to 
determine the physical scale. The uncertainty of this scale, at least in quenched 
calculations, is usually expected to be on the order of 10\%. 

As discussed in section \ref{sec:numerics} at least part of the deviations of 
(\ref{fpi}) from (\ref{fpilattice}) can be attributed to the absolute value 
approximation. However, it is by no means clear, whether a calculation in the 
complex plane will bring $f_\pi$ all the way up to the physical value. An 
indication that this may well {\it not} be the case is a comparison with $f_\pi$ 
from a modified Pagels-Stokar approximation~\cite{Pagels:hd,Roberts:1994dr}
\beqa
f_{\pi,PS}^2 &=& -\frac{N_c}{4 \pi^2} Z_2 \int dq^2 \: q^2 \: \frac{M(q^2) Z_f(q^2)}
{(q^2 + M^2(q^2))^2} \left(M(q^2) - \frac{q^2}{2} \frac{dM(q^2)}{dq^2}\right) \,.
\label{f-pi-eq}
\eeqa
This approximation is not affected by the absolute value problem, but in turn 
incorporates only the effects of the leading pion Bethe-Salpeter amplitude 
$E_{\pi}$ in the chiral limit. Thus the systematic errors are different. By 
comparison of (\ref{f-pi-eq}) with the full result in model 
calculations~\cite{Maris:1997tm,Alkofer:2002bp} one finds that the 
approximation (\ref{f-pi-eq}) should lead to an underestimation of the full 
$f_\pi$ of the order of 10\%. In our case we obtain 
\beq
f_{\pi,PS}^{N_f=2} = 65 \,\mbox{MeV} \hspace*{1cm} 
f_{\pi,PS}^{N_f=0} = 82 \,\mbox{MeV}. \label{PSfpi}
\eeq
Compared to (\ref{fpi}) we find a considerable increase in the unquenched 
value and no change for the quenched case. Within our quenched interaction we 
can also perform a full calculation in the complex plane using the techniques 
described in~\cite{Fischer:2005en}. We then find the value
\beq
f_{\pi,complex}^{N_f=0} = 89 \,\mbox{MeV}, \label{fpicomplex}
\eeq
which is indeed roughly 10\% larger than the result (\ref{fpi}) of the absolute 
value approximation and also 10\% larger than the result (\ref{PSfpi}) from the
Pagels-Stokar approximation. From this one would expect to find a value of
roughly $f_{\pi,complex}^{N_f=2} \approx 70-75 \,\mbox{MeV}$ from an unquenched
calculation in the complex plane. Clearly both, the calculated result 
(\ref{fpicomplex}) for the quenched case and the conjectured result for the 
unquenched case fall short of (\ref{fpilattice}) by almost twenty percent.  

A possible source of these shortcomings are errors in the overall scale of
our calculation. This scale was set by comparison with the lattice results 
for the gluon~\cite{Bowman:2004jm} and quark propagator~\cite{Bowman:2005vx}.
Indeed, already in~\cite{Bhagwat:2003vw} and again in~\cite{Fischer:2005nf} it 
was found that the pion decay constant calculated from an interaction fitted to 
quenched lattice data yields an underestimation of $f_\pi$ by about thirty percent.
This agrees with our findings above and supports the argument that there may be 
problems with the lattice scale. We also note that a rescaling of the lattice units 
would lead to larger constituent quark masses and in addition reduce the effect of 
the Goldstone contribution in the selfenergy. This would improve the agreement of 
the results shown in Fig.~(\ref{fig:quark}).

\subsection{Gell--Mann-Oakes-Renner relation and chiral condensate}\label{sec:chiral}

\begin{figure}[t]
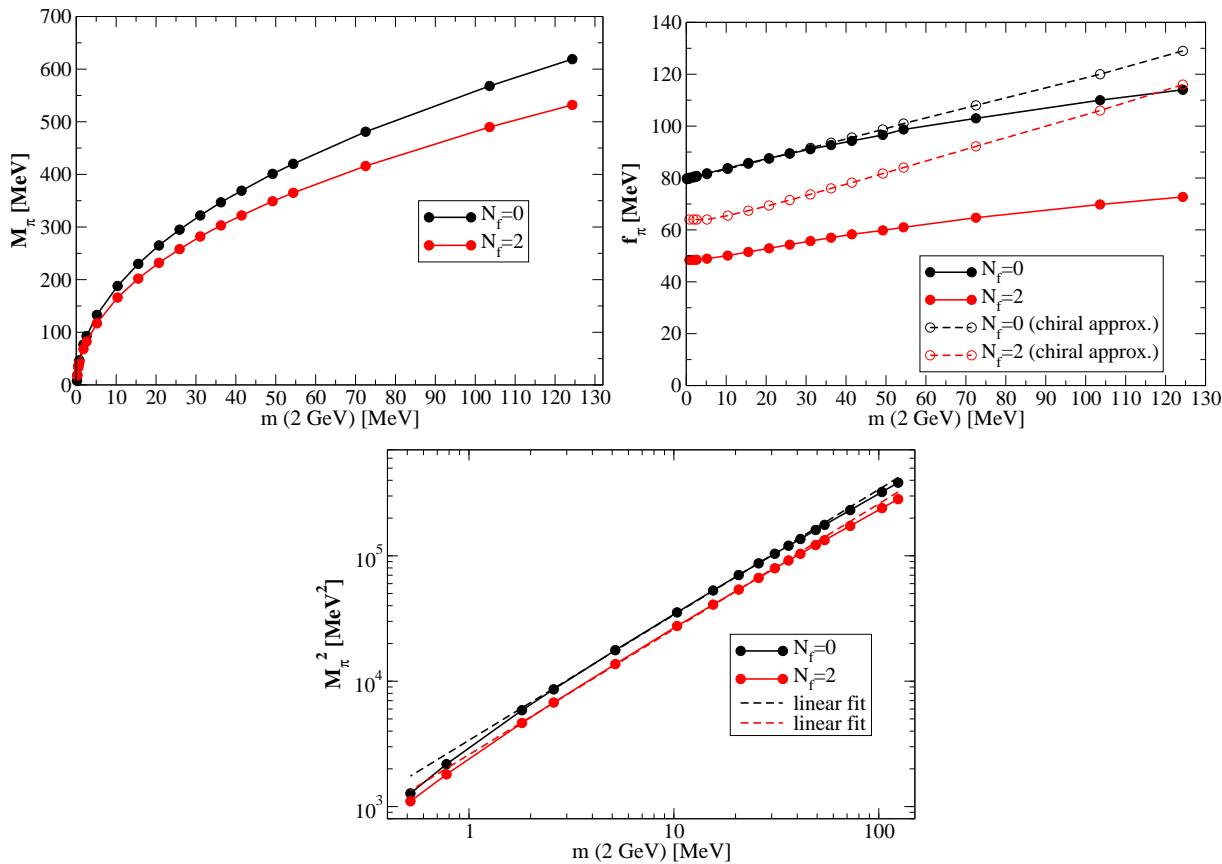

\centerline{\epsfig{file=Mpi_m.eps,width=8cm}
      \hfill\epsfig{file=fpi_m.eps,width=8cm}}
      \vspace*{2mm}
\centerline{\epsfig{file=GMOR.eps,width=8cm}}
\caption{The quenched and unquenched ($N_f=2$) pion mass and decay 
constant as a function of the renormalized quark mass $m(\mu)$ at $\mu=2$ GeV 
(MOM-scheme). The straight lines in the lower panel show the validity of
the Gell--Mann-Oakes-Renner relation.}
\label{fig:pimass}
\vspace{0mm}
\end{figure}

The Gell--Mann-Oakes-Renner relation 
\beq
M_\pi^2 \, \left(f_\pi^0\right)^2 = [m_u(\mu) + m_d(\mu)] \langle \overline{\Psi} 
\Psi \rangle^0 (\mu)
\eeq
can be derived from the inhomogeneous Bethe-Salpeter equations for the 
axialvector and pseudoscalar currents and the axialvector Ward-Takahashi 
identity~\cite{Maris:2003vk}. Since the left-hand side of this equation only 
contains physical observables, the dependence of the renormalization point of 
the current quark masses $m_u(\mu)$ and $m_d(\mu)$ and the chiral condensate 
$\langle \overline{\Psi} \Psi \rangle^0 (\mu)$ have to cancel each other. The 
individual dependence of the pion mass and the pion decay constant on the 
current quark mass is shown in the upper panel of Fig.~\ref{fig:pimass}. 
Clearly, one observes the characteristic square-root behavior of the pion mass
that becomes massless in the limit $m_{u/d} \rightarrow 0$. No qualitative 
changes can be seen by naked eye between the quenched and unquenched 
calculation. The situation is somewhat different for the pion decay constant. 
Here an almost linear behavior for sizable quark masses can be seen, with small 
deviations towards the chiral limit. Whether the different curvature for 
$m_{u/d} \rightarrow 0$ can be explained by the presence of chiral logs remains 
to be explored in a more refined calculation. In the lower panel of
Fig.~\ref{fig:pimass} we plot $M_\pi^2$ as a function of the current quark mass. 
Clearly, the dependence is linear for a wide range of current quark masses. The 
deviations from the linear fit for small quark masses may well be an
artifact of the increasing numerical uncertainties close to the chiral limit.
The deviations for larger quark masses show the onset of quantitative corrections 
to the Gell--Mann-Oakes-Renner relation. It is interesting to note, however, that 
these corrections play no role up to pion masses of $M_\pi = 400$ MeV. A similar 
result has been found in a recent lattice study using Wilson quarks and fine 
lattices~\cite{DelDebbio:2006cn}. 

The chiral condensate can be extracted directly from the quark propagator via
\beqa
-\langle \bar{\Psi}\Psi\rangle(\mu) := Z_2 \, Z_m \, N_c \,\mathrm{Tr}_D \int
\frac{d^4q}{(2\pi)^4} S_{0}(q^2) \,,
\label{ch-cond}
\eeqa
where the trace is over Dirac indices, $S_{0}$ is the quark propagator in the 
chiral limit and $Z_m$ is the quark mass renormalization factor that can also 
be determined from the quark-SDE. We obtain
\beq
\langle \overline{\Psi} \Psi \rangle^{0,N_f=2}_{\overline{MS}}(2 \, \mbox{GeV}) 
= (-240 \,\mbox{MeV})^{3}
\hspace*{1cm}
\langle \overline{\Psi} \Psi \rangle^{0,N_f=0}_{\overline{MS}}(2 \, \mbox{GeV}) 
= (-283 \,\mbox{MeV})^{3},
\eeq
which shows a sizable unquenching effect in the condensate. Whether this effect 
stays as large when the absolute value approximation is given up remains to be 
seen. Nevertheless it is instructive to use these values as input into the 
Gell--Mann-Oakes-Renner relation. This allows us to extract the pion decay 
constant once more from an independent source. We obtain
\beq
f_{\pi,GMOR}^{N_f=2} = 74 (3) \,\mbox{MeV} \hspace*{1cm} 
f_{\pi,GMOR}^{N_f=0} = 83 (3) \,\mbox{MeV}, 
\label{GMORfpi}
\eeq
for the absolute value approximation and 
\beq
f_{\pi,GMOR,complex}^{N_f=0} = 85 (3) \,\mbox{MeV}, \label{GMORfpicomplex}
\eeq
for the quenched calculation using complex momenta. The error subsumes numerical 
errors as well as errors in fitting $M_\pi^2(m)$, which has been performed by a 
least squares fit in the region from 
$m(2 \,\mbox{GeV}) = 3-40 \,\mbox{MeV}$.  By comparison with (\ref{fpicomplex}) 
we note that the GMOR-relation is satisfied at the 4 percent level for the 
central value in (\ref{GMORfpicomplex}). The difference of the corresponding 
value for the unquenched case and the result Eq.~(\ref{fpi}) shows once more that 
the absolute value approximation leads to a large error in the unquenched $f_\pi$.   

\subsection{Analytical properties of the quark propagator}\label{sec:analytical}

\begin{figure}[t]
\centerline{\epsfig{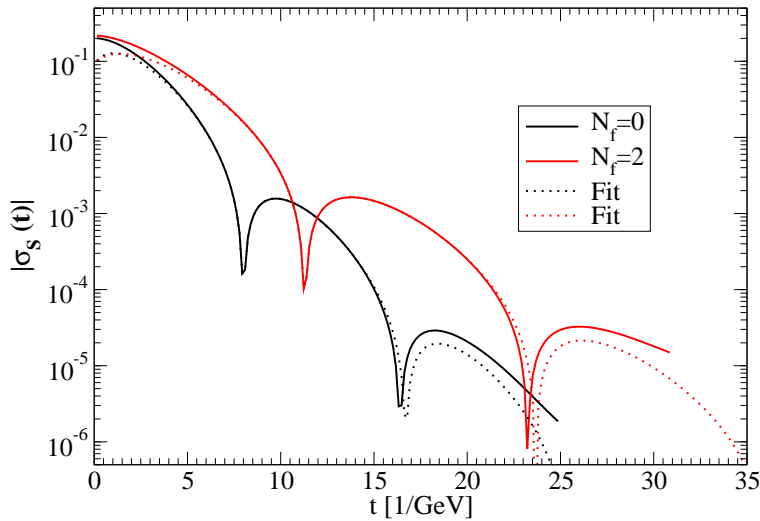}}
\caption{The absolute value of the Schwinger function $\sigma_S(t)$ of the 
quark propagator.}
\label{fig:analytic}
\end{figure}
\begin{figure}[t]
\label{test2}
\vspace{0mm}
\end{figure}

Finally we take a look at the analytic structure of the quark propagator in the 
complex momentum plane. We wish to stress that the results of this section have 
to be treated with caution, since the absolute value approximation is likely not 
to be reliable in this respect. Nevertheless, it may be instructive to have a 
first glance in the complex plane. To this end we determine the Schwinger-function
\beq
\sigma_S(t)\; =\; \int d^3x \int \frac{d^4p}{(2\pi)^4} \exp(i p \cdot x) 
\sigma_{S}(p^2),
\eeq
where $\sigma_{S}(p^2) = B(p^2)/(p^2 A(p^2)^2 + B(p^2)^2)$ is the scalar part of 
the dressed quark propagator. (This method has a long history, see 
{\it {\it e.g.}}~\cite{Alkofer:2003jj} and references therein). According to the 
Osterwalder-Schrader axioms of Euclidean field theory~\cite{Osterwalder:1973dx}, 
this function has to be positive to allow for asymptotic quark states in the 
physical sector of the state space of QCD. Conversely, positivity violations in 
the Schwinger function show that the corresponding asymptotic states (if 
present) belong to the unphysical part of the state space. In
Fig.~\ref{fig:analytic} we show results for the Schwinger function of
the quenched and unquenched quark propagator for the physical up/down quark of
Fig.~\ref{fig:quark}. Both functions are qualitatively similar. An excellent fit 
to the Schwinger function is obtained using the form~\cite{Stingl:1996nk}
\beq
|\sigma_S(t)|\; =\; |\, b_0 \exp(-b_1 t) \cos(b_2 t+b_3)\,| \quad , \label{cc}
\eeq
which corresponds to a pair of complex conjugate poles of the propagator in the 
time-like momentum plane. These poles correspond to a \lq quark mass' given by
$m = b_1 \pm i b_2$. In our quenched case this amounts to 
$m = 505(10) \pm i \, 360(10)$ MeV; the unquenched quark mass is 
$m = 351(10) \pm i \, 254(10)$ MeV. The mismatch of the fit with $\sigma(t)$ at 
small times shows the presence of additional analytic structure either in form 
of a cut along the real time-like momentum axis or in form of additional 
singularities further away from the origin. If this behavior is stable also 
when the absolute value approximation is overcome it contradicts the Gribov scenario 
given in~\cite{Gribov:1992tr}. However, as emphasized already above, definite 
conclusions can only be drawn from a complete calculation including the full 
complex momentum structure involved in the coupled system of the quark-SDE and 
the pion-BSE.

\section{Summary and outlook}\label{sec:sum}

In this work we investigated the pion back reaction on the quark propagator and 
resulting pion properties. To the best of our knowledge this is the first 
calculation of this kind in a non-perturbative continuum approach to QCD. We 
isolated the pion contribution to the quark-gluon vertex and identified an 
approximation to the quark-SDE which allows for the construction of a Bethe-Salpeter 
kernel in agreement with the axial Ward-Takahashi identity. This setup is powerful 
enough to analytically verify Coleman's theorem in 1+1 dimensions: chiral symmetry 
cannot be broken spontaneously there. Our main interest, however, is QCD$_{3+1}$, 
where we evaluated the back reaction effects numerically. Here we had to resort to 
an absolute value approximation, which is reliable as concerns the pion mass but 
problematic for the pion decay constant.

As a result we have obtained considerable unquenching effects in the quark mass 
function at small momenta. The screening effect of the pion interaction reduces
the quark mass $M(0)$ from $M^{N_f=0}(0)=477$ MeV to $M^{N_f=0}(0)=371$ MeV.
This effect becomes smaller for larger momenta and vanishes above $p=3$ GeV in
correspondence with the width of the Bethe-Salpeter amplitude of the pion.
Almost no corresponding effect in the quark wave function is seen.

The screening effect of the pion interaction also reduces the value of the
chiral condensate. The Gell--Mann-Oakes-Renner relation is satisfied with an
accuracy less than 4 \%. We obtain an almost linear relation between the 
squared pion mass and the current quark mass for pion masses up to 400 MeV
with only slight deviations up to 500 MeV. An unsolved problem is posed by
our low value of the pion decay constant. Certainly, part of the 
problem is the absolute value approximation mainly used in this work.
However, we also argued that the overall scale of our calculation, which is
based on the lattice results of~\cite{Bowman:2004jm,Bowman:2005vx}, may have 
to be corrected by about 20 \%.

Our findings are certainly reliable on a qualitative basis. To also obtain
reliable quantitative results we need to go beyond the absolute value 
approximation for the arguments in the internal quark propagators and pion 
amplitudes. Work in this direction is well under way. We then hope to better
understand the problem of the very low pion decay constant obtained in this 
work. Also, working in the complex momentum plane is mandatory to investigate
possible unquenching effects in the analytic structure of the quark propagator 
conjectured by Gribov. Finally, a complex treatment of the system is mandatory 
to explore decay thresholds in vector, axialvector and the scalar
meson sectors. An investigation of this mesons should be simplified by
use of the Maximum Entropy Method~\cite{Nickel:2006mm}.

\acknowledgments
We are grateful to Patrick Bowman for making the lattice data of 
ref.~\cite{Bowman:2005vx} available. This work has been supported by the 
Helmholtz-University Young Investigator Grant VH-NG-332.

\begin{appendix}
\section{Multiplicative Renormalizability of SDE and BSE \label{MR}}
Here we show explicitly that our approximation scheme for the Schwinger-Dyson 
and Bethe-Salpeter equations preserves multiplicative renormalizability (MR). 
To this end we rewrite (\ref{quarkdse2}) and (\ref{eq:bse2}) in symbolic 
notation and make explicit the dependence of the various quantities on the 
renormalization point $\mu$. We begin with the SDE:
\beqa
S^{-1}(p,\mu^2) &=& Z_2(\mu^2) \,S^{-1}_0(p) + g^2(\mu^2)\, Z_{1F}(\mu^2) \,
\int   S(q,\mu^2)\, Z(k^2,\mu^2) \,\Gamma_{YM}(k^2,\mu^2) \,L_{YM}(p,q,k) \nonumber\\
&&- 3 \int \left[      
      \Gamma^{\pi}(p,q,\mu^2) \, S(q,\mu^2)\,
      \Gamma^{\pi}(p,q,\mu^2) \,L_{\pi}(p,q,k) \right], \label{quark-mr}
\eeqa
where the functions $L_{YM}$ and $L_{\pi}$ subsume all $\mu$-independent 
quantities in this equation. Since 
\beq
S^{-1}(p,\mu^2) = i \pslash A(p^2,\mu^2) + B(p^2,\mu^2)
\eeq
and we have the relations
\beqa
A(p^2,\mu^2) &=& Z_2(\mu^2,\Lambda^2) A_0(p^2,\Lambda^2) \nonumber\\
B(p^2,\mu^2) &=& Z_2(\mu^2,\Lambda^2) B_0(p^2,\Lambda^2)
\eeqa
between the renormalized quantities $A,B$ and the unrenormalized, cutoff 
($\Lambda$) dependent quantities $A_0,B_0$, we note that the $\mu$-dependence of
the left hand side of Eq.~(\ref{quark-mr}) is given by $Z_2(\mu^2)$.  This is 
also trivially true for the tree-level term $Z_2(\mu^2) \,S^{-1}_0(p)$. It 
follows then that the whole equation is multiplicatively renormalizable if and 
only if each term of the self energy is also proportional to $Z_2(\mu^2)$. To 
show this we need the relations
\beqa
Z(k^2,\mu^2) Z_3(\mu^2,\Lambda^2) &=& Z_0(k^2,\Lambda^2) \nonumber\\
G(k^2,\mu^2) \widetilde{Z}_3(\mu^2,\Lambda^2) &=& G_0(k^2,\Lambda^2)\nonumber\\
g(\mu^2) Z_g(\mu^2,\Lambda^2) &=& g_0(\Lambda^2)  \label{zgren}
\eeqa
between the renormalized gluon and ghost dressing functions $Z,G$ and their
unrenormalized counterparts $Z_0,G_0$ and a similar relation for the coupling 
$g$. With the help of (\ref{zgren}) we extract the renormalization point 
dependence of the Yang-Mills interaction from Eqs.(\ref{v2})-(\ref{v5}):
\beq
\Gamma_{YM}(k^2,\mu^2) \sim \frac{Z_2(\mu^2)}{\widetilde{Z}_3(\mu^2)}.
\eeq
The renormalization point dependence of the Yang-Mills part of the self energy
is then given by
\beq
\Sigma_{YM}(p^2,\mu^2) \sim \frac{Z_{1F}(\mu^2)}{Z_g^2(\mu^2)
  \widetilde{Z}_3(\mu^2) Z_3(\mu^2)}.
\eeq
Taking into account the Slavnov-Taylor identities
\beq
Z_{1F} = Z_g Z_2 Z_3^{1/2}, \hspace*{1cm} \widetilde{Z}_1 = Z_g 
\widetilde{Z}_3 Z_3^{1/2}
\eeq
we arrive at
\beq
\Sigma_{YM}(p^2,\mu^2) \sim \frac{Z_2(\mu^2)}{\widetilde{Z}_1(\mu^2)},
\eeq
which contains the ghost-gluon vertex renormalization constant $\widetilde{Z}_1$. 
In Landau gauge the ghost-gluon vertex is finite and consequently this quantity 
can always be chosen to equal one by a suitable renormalization 
condition\footnote{In fact we do not even have to specify such a condition. The 
gluon propagator (\ref{fullfit}) used in our calculation has been calculated in 
refs.~\cite{Fischer:2002hn,Fischer:2003rp} with a bare ghost-gluon vertex. This 
choice is well motivated and enforces $\widetilde{Z}_1=1$}. Thus we arrive at the 
desired result $\Sigma_{YM}(p^2,\mu^2) \sim Z_2(\mu^2)$.

The other contribution to the quark self energy trivially gives the same result 
provided 
\beq
\Gamma^{\pi}(p,q,\mu^2) \sim Z_2(\mu^2).
\eeq
This dependence agrees with the one required in the norm integral, 
Eq.~(\ref{norm}), and the expression for $f_\pi$, Eq.~(\ref{eq:fpi}) and 
therefore completes the proof of MR for the SDE.

In the same symbolic notation the Bethe-Salpeter equation for the pion can be 
written as
\beqa
\Gamma^{\pi}(p,P) &=& g^2(\mu^2) \int \Gamma^{\pi}(q,P,\mu^2) \,S(q_-,\mu^2) 
                       \,S(q_+,\mu^2) \,Z_{1F} \,Z(k^2,\mu^2) \,
		       \Gamma_{YM}(k^2,\mu^2) \, J_{YM}(p,q,P) \nonumber\\
		    &&+  \int \Gamma^{\pi}(q,P,\mu^2) \,S(q_-,\mu^2) 
		      \,S(q_+,\mu^2) \,[\Gamma^{\pi}(q,p,P,\mu^2)]^2 
		      \,J_{\pi}(p,q,P)
\eeqa
where the functions $J_{YM}$ and $J_{\pi}$ again subsume all $\mu$-independent 
quantities. With the help of the relations given above it is easy to verify
that all terms on the left and right hand side of this equation are proportional 
to $Z_2$, which leaves the equation multiplicatively renormalizable.

\end{appendix}


\end{document}